\documentclass{article}

\usepackage{graphicx,amsmath, amsfonts}
\usepackage[english]{babel}
\usepackage[utf8]{inputenc}
\usepackage[T1]{fontenc}
\usepackage{color}

\newtheorem{numer}{\hspace{-4mm}($\spadesuit$\hspace{-1mm}}

\newcommand{\popr}[1]{
 { #1} 
}

\newtheorem{tnumer}{\hspace{-1mm}($\clubsuit$\hspace{-1mm}}
\newcommand{\trefle}[2]{
\begin{tnumer}
\hspace{-1.5mm}{\em )}
\label{#1} 
#2 
\end{tnumer}
}
\newcommand{\rtrefle}[1]{($\clubsuit$ \hspace{-1.5mm} \ref{#1})}

\newcommand{\dd}{{\mathbb D}}

\newcommand{\mlodszy}[1]{ {\rightarrow}_{#1}}

\newcommand{\eop}{\hfill {$\Box$}}
\def\bi{\begin{itemize}}
 \def\ei{\end{itemize}}
 
\newcommand{\calp}{{\cal P}}
\newcommand{\calpbu}{{\cal P}_{\beta_{wu}}}

\newtheorem{theorem}{\hspace{-.5mm}Theorem}
 \newtheorem{prop}[theorem]{\hspace{-.5mm}Property}
 \newtheorem{lemmaa}[theorem]{\hspace{-.5mm}Lemma}
 
 \newtheorem{definitionn}[theorem]{\hspace{-.5mm}Definition}
\newtheorem{observ}[theorem]{\hspace{-.5mm}Observation}
\newtheorem{remark}[theorem]{\hspace{-.5mm}Remark}

\newcommand{\propertyyy}[1]{\vspace{1.2mm}\begin{prop} #1 \end{prop} \vspace{1.2mm}}
\newcommand{\theoremmm}[1]{\vspace{1.5mm}\begin{theorem} #1 \end{theorem} \vspace{2mm}}
\newcommand{\lemmaaa}[1]{\vspace{0.5mm}{\em \begin{lemmaa} #1 \end{lemmaa} }\vspace{0.5mm}}

\newcommand{\definitionnn}[1]{\vspace{1.1mm} {\em \begin{definitionn} #1 \end{definitionn} } \vspace{1.4mm}}

\newcommand{\observationnn}[1]{\vspace{1.2mm}\begin{observ} #1 \end{observ} \vspace{1.4mm}}

\newcommand{\citep}[1]{\cite{#1}}

\author{Tomasz Gogacz,Jerzy Marcinkowski}

\title{Converging to the Chase -- a Tool for Finite Controllability 
\footnote{This is the full version of an extended abstract published in the LICS 2013 proceedings}}

\begin{document}

\maketitle 

\begin{abstract}

We solve a problem, stated in \citep{CGP10},  showing that Sticky  Datalog$^\exists$, 
defined in the cited paper as an element of the Datalog$^\pm$ project, has the Finite 
Controllability property. 
In order to do that, we develop a technique, which we believe can have further applications,
of approximating 
 $Chase({\cal T},\dd)$, for a database instance $\dd$ and a set of tuple generating dependencies and Datalog rules $\cal T$,
 by an infinite sequence of finite structures, all of them being models of $\cal T$ and $\dd$.
 \end{abstract}

%


\section{Introduction\\}

Tuple generating dependencies (TGDs), recently also known as Datalog$^\exists$ rules, are studied in
various areas, from database theory to description logics and in various contexts. The context we are 
interested in here is  computing certain answers 
to queries in the situation when  some semantical information about the database is known (in the form of theory $\cal T$, consisting of TGDs),  but the knowledge of the database facts is limited, so that the known set of facts $\dd$ does not necessarily satisfy the dependencies of $\cal T$.

It is easy to see that query answering in presence of TGDs is undecidable. As usually in such situations 
many sorts of syntactic restrictions on the dependencies are considered, which imply decidability while keeping as much 
expressive power as possible. Recent new interest in such restricted logics comes from the Datalog$^\pm$ project,
led by Georg Gottlob, whose aim is translating  concepts and proof techniques 
from database theory to description logics and {\em bridging an apparent gap in expressive power between database query languages and description logics (DLs) as ontology languages, extending the well-known Datalog language in order
to embed DLs} \citep{CGL09}. 

From the point of view of Datalog$^\pm$ and of this paper, the interesting logics are:

\noindent
{\bf Linear Datalog$^\exists$} programs. They consist of TGDs which, as the body, have a single atomic formula, 
and this formula is joinless -- each variable in the body occurs there only once. 
Let us note that allowing variable repetitions in the heads does not change the Finite Controllability status of a program, as 
we can always remember the equalities as part of the relation name, so 
we  w.l.o.g. assume that such repetitions are not allowed (see  Section \ref{porzadeknabiurku} for much
more about this issue).
The {\bf Joinless Logic} we consider in this paper is a  generalization of Linear Datalog$^\exists$, in the sense that 
we no longer restrict the body of the rule to be a single atom, but we still demand that each variable occurs in the body only once\footnote{The term ''Joinless Logic'' was used in [CGP10a] (Theorem B.2 there) to denote logic which is not really joinless -- a variable may occur more than once there, but only in one atom in the body. It is however very easy to see that any TGD can be simulated by one TGD and one Datalog rule, which are ''joinless'' in this sense. Unlike [CGP10a], when we say ''joinless''  we really mean ''joinless''.} 

 \vspace{1mm}

\noindent
{\bf Guarded Datalog$^\exists$} is an extension of Linear Datalog$^\exists$. 
 A TGD is guarded if it has an atom, in the body, containing all the variables 
that occur anywhere else in the body. Clearly, Linear Datalog$^\exists$ programs are guarded, as they only have one atom in the body.\vspace{1mm}

\noindent
{\bf Sticky Datalog$^\exists$} is a logic introduced in \citep{CGP10} and then extended in \citep{CGP10+/-} as 
Sticky-Join Datalog$^\exists$. Theory $\cal T$  is Sticky, if some positions in the predicates from the signature of $\cal T$ 
 can be marked as ''immortal'' in such a way that the following conditions are satisfied:
\begin{itemize}
\item If some variable occurs \popr{in at least one} immortal position in the body of a rule from $\cal T$ then the same 
variable must occur in an immortal position in the atom being the head of the same rule. 

\item If some variable occurs more than once in the body of a rule from $\cal T$ then this
variable must occur in an immortal position in the atom being the head of the same rule. 
\end{itemize}

The above  definition of Sticky Datalog$^\exists$ is a slightly different 
wording\footnote{
Both the definitions of  Sticky  Datalog$^\exists$ involve comparing the set 
$J$ of positions where joins occur with the set $V$ of positions where variables are allowed to vanish. 
In \citep{CGP10} authors have chosen to state the condition in the language of pullback of $V$ by by the rules of $\cal T$ while
 we prefer to think in terms of pushforward of $J$ by the rules of $\cal T$.}
 of (equivalent, when restricted to single-head TGDs) Definition 1 from \citep{CGP10}, and 
resembles what in the paper \citep{CGP10+/-}  is  called
''the sticky-join property'' (see Section 5.1  in \citep{CGP10+/-}). Actually, both
Theorem \ref{mainwniosek} of our paper and its proof hold for any  possible logic having the sticky-join property, which includes  
Sticky Datalog$^\exists$  and Sticky-Join Datalog$^\exists$ (which is a version defined in  \citep{CGP10+/-}).
In fact, the difference between the two logics can only be seen if repeated variables in the heads of the rules are 
allowed  and, as we said before, from the point of view of Finite Controllability we can disallow them w.l.o.g..\vspace{2mm}

Apart from decidability, the  properties of such logics which are  considered desirable
and receive a lot of attention are:\vspace{1mm}

\noindent
{\bf Bounded Derivation Depth property (BDD).} A set $\cal T$ of TGDs has the  bounded derivation depth property if for each UCQ $\Psi$  
there is a constant $k_\Psi\in \mathbb N$, such that for each database instance  $\dd$
if $Chase({\cal T},\dd)\models \Psi$ then $Chase^{k_\Psi}({\cal T},\dd)\models \Psi$.
The BDD property turns out to  be equivalent to positive existential first order rewriteability : 

\begin{theorem}
$\cal T$ has the BDD property if and only if
for each UCQ $\Psi$ there exist a UCQ $\Phi$ such that for each database instance $\dd$ (finite or not) 
it holds that
$Chase({\cal T},\dd)\models \Psi$ if and only if $ \dd\models \Phi$.
\end{theorem}

This theorem is stated in \citep{CGL09}, not as an equivalence however, but only as the  ''only if'' implication -- if a theory  is BDD then queries are  rewritable as UCQs.
We believe that the proof of the 
''if'' implication is folklore, but let us include it here, for sake of completeness: 

Fix a theory ${\cal T}$ and  assume that query $\Psi$ is rewritable.
Let $\Phi=$ $\phi_1 \vee \phi_2\ldots \vee \phi_m$  be the rewriting, where each $\phi_i$ is a conjunctive query. For each $\phi_i$
let $M_i$ be the canonical structure of $\phi_i$. Clearly, for each $i$ we have $M_i\models \Phi$ so also for each $i$ there is 
$Chase({\cal T},M_i)\models \Psi$. Let $k_i$ be a natural number such that   $Chase^{k_i}({\cal T},M_i)\models \Psi$.
Now define $k_\Psi$ as $\max\{ k_i: 1\leq i\leq m\}$. It is now easy to see that, for any $\dd$, it holds that
if $Chase({\cal T},\dd)\models \Psi$ then $Chase^{k_\Psi}({\cal T},\dd)\models \Psi$.\eop
\vspace{2mm}

\noindent
{\bf Finite Controllability (FC).} A set $\cal T$ of TGDs has the 
finite controllability property if for each UCQ  $\Psi$ and each database instance $\dd$ 
if  $Chase({\cal T},\dd) \not\models \Psi$ then
there exists a finite structure $M$ such that $M\models {\cal T},\dd$ but $M\not\models\Psi$. \vspace{1mm}

A logic is said to be FC (or BDD)  if each $\cal T$ in this logic is FC (BDD). 
A triple $\cal T$, $\dd$, $\Psi$ such that $Chase({\cal T},\dd) \not\models \Psi$ but for each finite 
structure $M$ if  $M\models {\cal T},\dd$ then also $M\models \Psi$ will be called a {\bf counterexample for FC}.
It  is usually quite easy to see whether a given logic is BDD and 
 it is usually very hard to see whether it is FC.\vspace{2mm}

\noindent
{\bf Previous works.}
The query answering problem for Linear Datalog$^\exists$ (or rather for Inclusion Dependencies, 
which happens to be the same notion as Linear Datalog$^\exists$) was shown to be decidable 
(and PSPACE-complete) in \citep{JK84}. The problem which was left open in \citep{JK84} was finite 
controllability -- since we mainly consider finite databases, we are not quite happy with the answer
that ''yes, there exists a database $\bar \dd$, such that $\bar \dd\models {\cal T},\dd,\neg \Psi$''
if all counterexamples $\bar \dd$ for $\Psi$ we can produce  are infinite. 
This problem was solved by Rosati \citep{R06}, who proved, by a complicated argument, that IDs (Linear Datalog$^\exists$) 
have the finite controllability property. His
result was improved in \citep{BGO10} where FC is shown for Guarded Datalog$^\exists$. 

Sticky Datalog$^\exists$ was introduced in \citep{CGP10}, where it was also shown to have the BDD property and where 
the question of the FC property of this logic was stated as an open problem. 
The argument, given in \citep{CGP10}, motivating the study of Sticky Datalog$^\exists$
is that it {\em can express assertions having  compositions of roles in the body,
which are inherently non-guarded. 
 Sticky sets of TGDs can  express constraints and rules
involving joins. We are convinced that the overwhelming
number of real-life situations involving such constraints can
be effectively modeled by sticky sets of TGDs. Of course,
since query-answering with TGDs involving joins is undecidable in general, we somehow needed to restrict the interaction of TGDs, when joins are used. But we believe that the
restriction imposed by stickiness is a very mild one. Only
rather contorted TGDs that seem not to occur too often in
real life violate it. For example, each singleton multivalued
dependency (MVD) is sticky, as are many realistic sets of
MVDs }\citep{CGP10}.\vspace{2mm}

\noindent
{\bf Our contribution.} 
We show two finite controllability results. Probably the more important of them is:

\theoremmm{
\label{mainwniosek}
Sticky  Datalog$^\exists$  is FC.
}

\noindent
But this is merely a corollary to a theorem that we consider the main technical achievement of this paper:

\theoremmm{
\label{maintheorem}
Joinless Logic  is FC.
}

To prove Theorem \ref{maintheorem} we propose a technique, which we think is quite elegant\footnote{This is just our opinion. The reader has of course the right to have his own.}, and relies on two main ideas.
One is that we carefully trace the relations (we call them ''family patterns'') between pairs of elements of $Chase$ which are 
ever involved in one atom. The second idea is to consider  an infinite sequence of equivalence relations, defined by
the types of families  which the elements (and their ancestors) are members of, and construct an infinite sequence of models
as the quotient structures of these equivalence relations. This leads to a sequence of finite models, that, in a sense, ''converges'' to  $Chase$.

What concerns  the Joinless Logic as such, we prefer not to make exaggerated claims about its importance.
We see it  just as a mathematical tool --
the $Chase$ resulting from a Joinless  theory is a huge and very complicated structure, much more complex than 
the bounded tree-width $Chase$ resulting from guarded (or Linear) TGDs,  and the ability to control it can give
insight into chases generated by logics enjoying better practical motivation -- 
Theorem \ref{mainwniosek} serves here as a good example. But still Theorem \ref{maintheorem}
 is a very strong generalization of the result of Rosati about Linear Datalog$^\exists$,
which itself was viewed as well motivated,
while the technique we develop in order to prove it is powerful enough
to give, as a by-product, an easier proof of the finite controllability  result for sets of 
guarded TGDs \citep{BGO10}.
It also appears that rules with Cartesian products, even joinless, can be seen as interesting  from some sort of  practical 
point of view, motivated by Description Logics (where they would be called ''concept products''). After all, ''All Elephants are Bigger than All Mice'' \citep{RKH08}.\vspace{1.5mm}

\noindent
{\bf Open problem: BDD/FC conjecture.}
Does the BDD property always imply FC? In the proof of Theorem \ref{mainwniosek} we do not seem to use much much  more than just Theorem \ref{maintheorem} and 
the fact that Sticky  Datalog$^\exists$ is BDD. In our parallel paper \citep{GM13} we show  that each theory
over a binary signature which is BDD is also FC. We also explain there why the full conjecture is not so easy to prove. \vspace{1.5mm}

\noindent
{\bf Outline of the paper.}
Next section is devoted to preliminaries. Basic concepts are explained there and 
notations are introduced. 

In Section \ref{nastepna} we prove Theorem \ref{mainwniosek},
assuming Theorem \ref{maintheorem}. 

The proof of Theorem \ref{maintheorem}, which is the main technical
contribution of this paper, is presented in Sections \ref{acontrario}--\ref{ch}.

\vspace{-2mm}


\section{Preliminaries}\label{preliminaries}

Most of the notions and notations in this paper are standard for mathematical logic and database theory. In particular, if $\phi$ is a formula,
$\Psi$ is a set of formulas and $\cal M$ is a structure, then by ${\cal M}\models \phi$ we mean that $\phi$ is true in $\cal M$ and by
${\cal M}\models \Psi$ we mean that each formula in $\Psi$ is true in $\cal M$. By $\Psi\models \phi$ we mean that for each structure $\cal M$ such that
${\cal M}\models \Psi$ there is also ${\cal M}\models \phi$.

Let us remind the reader that  a {\em tuple generating dependency} (TGD), or a Datalog$^\exists$ {\em rule} (or just ''rule'') 
 is a formula of the form 
$$\forall \bar x \; (\Phi(\bar x) \Rightarrow \exists y \; Q(y,\bar y))$$
 where 
$\Phi $ is a \popr{conjunction of atoms} (a conjunctive query  without existential quantifiers), $Q$ is a relation symbol, $\bar x, \bar y$ are tuples of variables and
$\bar y\subseteq \bar x$. The universal quantifier in front of the formula is usually omitted. Notice that w.l.o.g we only consider single-head TGD, which means that there 
is always only one atom in the head (i.e. right hand side) of a rule.
By a {\em theory} we mean a finite set consisting of some TGDs and some Datalog rules (which are TGDs without the existential quantifier in the head). 

For a theory $\cal T$
and  a database instance $\dd$ the structure $Chase({\cal T}, \dd)$ is defined in the  standard way, and by 
$Chase^i({\cal T}, \dd)$ we mean the structure being the $i$-th stage of the fixpoint procedure leading to $Chase({\cal T}, \dd)$.

More precisely,  we define $Chase^0({\cal T}, \dd)=\dd$. Once 
$Chase^i({\cal T}, \dd)$ is defined, we define $Chase^{i+1}({\cal T}, \dd)$ as the superstructure of $Chase^i({\cal T}, \dd)$ being 
the result of the following 
procedure:\vspace{2mm}


{\bf for each} rule $\Phi(\bar x) \Rightarrow \exists y \; Q(y,\bar y)$ from $\cal T$ and {\bf for each} 
valuation $\rho$ mapping variables in $\bar x$ to elements of  $Chase^{i}({\cal T}, \dd)$ 
 such that $Chase^{i}({\cal T}, \dd)\models \Phi(\rho(\bar x))$ but there is 
no  $b$ such that $Chase^{i}({\cal T}, \dd)\models  Q(b,\rho(\bar y))$, we add to  $Chase^{i+1}({\cal T}, \dd)$ a new element $b$ and
the atomic fact $  Q(b,\rho(\bar y))$;\vspace{1mm}

similarly, {\bf for each} Datalog rule from $\cal T$ and {\bf for each} relevant valuation an atomic fact is added to  $Chase^{i+1}({\cal T}, \dd)$ if it was not yet there.\vspace{2mm}

Then  $Chase({\cal T}, \dd)$  is defined as the union of all $Chase^i({\cal T}, \dd)$ for $i\in \mathbb N$. We often write  $Chase({\cal T})$ (or  $Chase$) instead of 
 $Chase({\cal T}, \dd)$ when $\dd$ (and $\cal T$) can be easily guessed from the context. Notice that when we say ''we add $b$ to $Chase^{i+1}({\cal T}, \dd)$'' we think of relational
structures as the mathematicians do -- as of a set of elements. But when we say ''add an atomic fact to  $Chase^{i+1}({\cal T}, \dd)$'' then we see a structure in the way
consistent with the database tradition -- as a set of facts. We will feel free to use both conventions, depending on which is more convenient at the moment.

Notice that the chase procedure as we define it above is  {\em standard} (lazy) chase. Unlike {\em oblivious chase}, which is also often 
considered in database theory, standard chase   adds new elements (and atoms which involve them) only when they are needed, that is when the body of some rule 
is satisfied for some valuation but the head of this rule is not. The choice of standard chase has an implication that will later be useful:
the standard chase procedure is idempotent, which means that 
 $Chase({\cal T}, \dd) =  Chase({\cal T}, Chase({\cal T}, \dd))$.

Clearly, we have $Chase({\cal T}, \dd)\models \dd, \cal T$, but there is no reason to think that $Chase^i({\cal T}, \dd)\models \cal T$
for any $i\in \mathbb N$.
Since $Chase({\cal T},\dd)$ is a ''free structure'',  it is well known, and very easy to see,  that
 for any query $\Phi$ 
 (being a union of positive conjunctive queries, or UCQ; remember that {\bf all queries we consider in this paper are positive}) 
  $\dd,{\cal T}\models \Phi$,
if and only if $Chase({\cal T}, \mathbb{D}) \models \Phi$.\vspace{1.5mm}

\noindent
{\bf A remark about notations.}  For any syntactic object $X$ by $Var(X)$ we will mean the set of all the variables in $X$. 

Letters $P$, $Q$ and $R$  will denote predicates or atoms of variables.
Letters $A,B,C, D$ will denote atoms of elements of $Chase$.
 $PP$ will be used for parenthood predicates (which are a special sort of predicates in our proof) and sometimes also for parenthood atoms. 
 
To denote elements of $Chase$ we will use $a,b,c,d$, while $i,j,k$ will 
 be positions in atoms or other small numbers. 

$F,G$ will be family orderings, and $\gamma$ and $\delta$ will be functions occurring in the family patterns -- something we develop
in Section \ref{families} and use extensively then. 

For an atom $B=  Q(b_1,b_2...b_k)$ (where $b_1,b_2...b_k$ are constants in $Chase$) we define a notation $B(i)= b_i$. The same applies for atoms of variables. 

The letter $M$ is always used to define a relational structure (usually a finite one). $\mathbb D$ is also used in this context, usually as the initial 
database instance on which chase is run.

$\Psi$ and  $\Phi$ are formulae, often  unions of conjunctive queries. The characters $\phi$, $\psi$ and $\beta$ are used to denote conjunctive queries (or just conjunctions 
of atoms). 

When we say ''conjunctive query'', or UCQ, we usually mean a boolean conjunctive query or boolean UCQ. This in particular applies (w.l.o.g.) to the definitions of FC and BDD. 
In order to keep the notation as light as possible, when talking about boolean CQs we often omit the existential quantifiers in front.


\section{From Joinless Logic to Sticky  Datalog$^\exists$}\label{nastepna}

This Section is devoted to the proof of Theorem \ref{mainwniosek} (assuming Theorem \ref{maintheorem}).

For a sticky theory $\cal T$ let  ${\cal T}_0$ be the subset of $\cal T$ that consists of all the joinless rules in $\cal T$.

A pair $\dd, \cal T$, where $\dd$ is a database instance, will be called weakly saturated if $\dd\models {\cal T}_0$. 
So  if $\dd, \cal T$ is weakly saturated then each  new element in $Chase(\dd, \cal T)$ must have some (sticky) 
join in its derivation and, in consequence each atom of $Chase({\cal T},\dd)$ is either an atom of $\dd$ or it 
contains some constant from $\dd$ in a marked position. This is 
because  (if  $\dd, \cal T$ is  weakly saturated) the only way for ${\cal T}$ to derive any atoms which are not in $\dd$ 
is to use some rule with the sticky join, which requires immortalizing one of the arguments

Suppose now that Sticky  Datalog$^\exists$ is not FC and we will consider counterexamples ${\cal T}, \dd, \Phi$ for FC with sticky $\cal T$.
By ''arity'' of $\cal T$ we will mean  the maximal arity of
atoms in the heads of the rules of $\cal T$.
We will call a counterexample ${\cal T}, \dd, \Phi$ {\em minimal} if the arity of $\cal T$ is smallest possible. 
  
We are  going to prove two Lemmas: 

\lemmaaa{\label{jedynka}
\popr{Suppose a triple ${\cal T}, \dd, \Phi$ is a minimal counterexample for FC. Then the the pair $\dd, \cal T$ is not weakly saturated.}
}

\lemmaaa{\label{dwojka}
\popr{Let $\cal T$, $\dd$, $\Phi$ be a counterexample for FC.}
There is a finite database instance $\dd'$ such that the pair $\dd', \cal T$ is weakly saturated and 
the triple  $\cal T$, $\dd'$, $\Phi$ is also a counterexample for FC. 
}

Notice that  proof of Theorem \ref{mainwniosek} will be finished 
once the two above lemmas are proved. This is because the assumption that a minimal counterexample exists will lead to a contradiction: 
by Lemma \ref{dwojka} we will be able to get a minimal weakly saturated counterexample -- something that is ruled out by Lemma \ref{jedynka}. 

 Theorem \ref{maintheorem} will be used to prove 
 Lemma \ref{dwojka}.\vspace{1mm}

\noindent
{\em Proof of Lemma \ref{jedynka}:} Let $\cal T$, $\dd$, $\Phi$ be a counterexample for FC, with $l$ being the arity of $\cal T$. 
Suppose the pair $\dd, \cal T$ was  weakly saturated.
We will construct a new sticky theory ${\cal T}_\dd$ of \popr{arity} at most $l-1$, over a new signature $\Sigma_\dd$ 
and a new query $\Phi_\dd$, such that the triple ${\cal T}_\dd$, $\emptyset$, $\Phi_\dd$ is also a 
counterexample for FC. This will contradict the assumption that $l$ was minimal possible, and thus end the proof of the Lemma.

Let us start from the definition of $\Sigma_\dd$. For a predicate $Q\in \Sigma$, of arity $j$, and 
for a  partial function $\gamma:\{1,2,\ldots j\}\rightarrow \dd$ let $Q_\gamma$ be a new 
predicate, of arity $j-|Dom(\gamma)|$. $\Sigma_\dd$ will be the set of all possible predicates $Q_\gamma$,
where $Q$ and $\gamma $ are as above. Since we did not assume that $\gamma$ is non-empty we have that 
$\Sigma \subseteq \Sigma_\dd$ (we identify $Q$ with $Q_\emptyset$).

To denote the predicates from $\Sigma_\dd$ we are going to use the notational convention that will now  be described 
 by an example. If $Q(\_,\_,\_)$ is a ternary predicate from $\Sigma$,  $\gamma=\{\langle 2, c\rangle\}$ and
 $\gamma'=\{\langle 1, c\rangle, \langle 3, a\rangle \}$ then $Q_\gamma$ will be denoted as  $Q(\_,c,\_)$ and
$Q_{\gamma'}$ will be denoted as  $Q(c,\_,a)$. Notice that the  $a$ and $c$ in $Q(\_,c,\_)$ and $Q(c,\_,a)$ are no longer
understood to be constants being arguments of the predicate. They are now part of the name of the predicate. 
Notice that  $|Dom(\gamma)|=1$ and indeed  $Q(\_,c,\_)$ is a binary relation, while  $|Dom(\gamma')|=2$ and $Q(c,\_,a)$ 
is a unary relation. 

As we are never going to use the constants from $\dd$ as arguments in atoms over relations from $\Sigma_\dd$, the above
notational convention does not lead to confusion as long as we only talk about atoms over $\Sigma_\dd$.  
But atoms over $\Sigma_\dd$  can easily be
confused with atoms over $\Sigma$ with constants from $\dd$ as arguments. And this confusion is exactly what we want!

If $\rho$ is a total function then $Q_\rho$ is an arity zero predicate. In particular each
atom of the database instance $\dd$ (over $\Sigma$) can be read as a zero arity predicate over $\Sigma_\dd$.

We are now going to define ${\cal T}_\dd$. 

For a rule $T$ from $\cal T$ by    a constantification\footnote{Our constantification trick is not claimed to be any sort of novelty -- see for example 
Constantification technique is by no means new. See for example the comment after Theorem 12.5.2 in \citep{AHV1995}
}
 of 
$T$ we will mean a formula $\sigma(T)$, where  $\sigma$ is 
a mapping that assigns constants from $\dd$ to some of the variables from $Var({T})$, in such a way that
for at least one variable $x\in Dom(\sigma)$ this $x$ appears in a marked position in $T$ 
(we mean here the marking of immortal positions, from the definition of Sticky Datalog$^\exists$). 
For example $Q(c,y,z)\Rightarrow \exists w \; Q(c,z,w)$ (where $c\in \dd$) is a  constantification of 
$Q(x,y,z) \Rightarrow \exists w \; Q(x,z,w)$ if position 1 is marked in $Q$.
Clearly, a constantification of a rule from $\cal T$ is (or ''can be seen as'') a rule over $\Sigma_\dd$.

Let now theory ${\cal T}_\dd$ over $\Sigma_\dd$ consist of 
 all the facts from $\dd$ (which now are, as we mentioned before, 
zero arity facts) and all the possible constantifications of  rules from $\cal T$. It is not hard to see that 
 ${\cal T}_\dd$  is also sticky (hint: mark as immortal the same positions as in $\cal T$), and that the \popr{arity} of 
${\cal T}_\dd$ is at most  $l-1$.

Let now $\cal C$ be the set of all atoms of $Chase({\cal T},\dd)$ (in the standard notation) and 
let  ${\cal C}_1$ be the set of all atoms of $Chase({\cal T}_\dd, \emptyset)$ (written using the above notational convention).

The assumption that  the pair $\dd, \cal T$ is  weakly saturated implies now:

\observationnn{ ${\cal C}={\cal C}_1$.}

For the {\bf proof of the Observation} recall that each atom of $Chase({\cal T},\dd)$ is either an atom of $\dd$ or it 
contains some constant from $\dd$ in a marked position. This is 
because  (as  $\dd, \cal T$ is  weakly saturated) the only way for ${\cal T}$ to derive any atoms which are not in $\dd$ 
is to use some rule with the sticky join, which requires immortalizing one of the arguments. And, when restricted to
atoms which contain some constant from $\dd$ in a marked position, 
the theories $\cal T$ and ${\cal T}_\dd$ derive exactly the same atoms.\eop

Now let us define $\Phi_\dd$ as the disjunction of all possible queries $\sigma(\Phi)$, 
where  $\sigma$ is 
a mapping that assigns constants from $\dd$ to some of the variables from $Var({\Phi})$ By distributivity,
if $\Phi$ was a UCQ then also  $\Phi_\dd$ is a UCQ. And
$Chase({\cal T},\dd)\models \Phi$ if and only if  $Chase({\cal T},\dd)\models \Phi_\dd $, which,
by the above Observation,  is equivalent to 
$Chase({\cal T}_\dd, \emptyset)\models \Phi_\dd $. Since we assumed that the triple $\cal T$, $\dd$, $\Phi$ is a counterexample for FC, this implies that $Chase({\cal T}_\dd, \emptyset)\not\models \Phi_\dd $. 

In order to prove that ${\cal T}_\dd$, $\emptyset$, $\Phi_\dd $ is a  counterexample for FC we still need to
show that for each finite structure $M$ over $\Sigma_\dd$ there is $M\models \Phi_\dd$. So suppose there was a finite 
$M$ such that $M\models {\cal T}_\dd$ and $M\not\models \Phi_\dd$. 
Define a new finite model $M^\dd$ as a structure over $\Sigma$, containing all the
elements of $M$ and all the elements of $\dd$, and all the atoms true in $M$. Of course the atoms true in $M$ were over
the signature $\Sigma_\dd$, but to define $M^\dd$ we  read them as atoms over $\Sigma$. It is easy to see that 
$M^\dd \models {\cal T}$ and $M^\dd \not\models {\Phi}$, which is however impossible as the triple  ${\cal T}$, $\dd$, $\Phi$ 
was a  counterexample for FC. \eop \vspace{2mm}


\noindent
{\em Proof of Lemma \ref{dwojka}:} Since Sticky  Datalog$^\exists$ enjoys the BDD property, we know that there exists 
a positive FO rewriting of $\Psi$, which is such 
a UCQ $\bar\Psi$  that for each database instance $\cal M$ (finite or not) it holds that ${\cal M}\models \bar\Psi$ if and only if
$Chase({\cal M},{\cal T})\models \Psi$.

Clearly, $Chase(Chase(\dd,{\cal T}_0), {\cal T})= Chase(\dd,{\cal T})$. So $Chase(\dd,{\cal T}_0) \not\models\bar\Psi\;$
 (as $Chase(\dd,{\cal T})\not\models \Psi$).

Since ${\cal T}_0$ is joinless, we know, from Theorem \ref{maintheorem}, that there exists a finite structure
$\dd'$ such that $\dd'\models {\cal T}_0, \dd$ but $\dd'\not\models \bar\Psi$. Notice that  the pair $\cal T$, $\dd'$ is weakly saturated.

Since $\dd'\not\models \bar\Psi$,
using again the fact that $\bar\Psi$ is the FO rewriting of $\Psi$,
 we get $Chase(\dd',{\cal T})\not\models \Psi$.  It remains to be shown that for each
finite structure $M$, if $M\models \dd',{\cal T}$ then $M\models \Psi$. But, since $\dd'\models \dd$,
the structure $M$ is a model of $\dd$ and we assumed that $M\models \Psi$ holds for each finite model of $\dd$ and $\cal T$.\eop

\section{Assumption a contrario and the structure of the proof}\label{acontrario}

Sections \ref{acontrario} -- \ref{ch} are devoted to the proof of Theorem  \ref{maintheorem}.\medskip

It is an a contrario proof so {\bf we assume now that} there exists a counterexample ${\cal T}_C$, $\dd_C$, $\Phi_C$ for FC, with
${\cal T}_C$ being a joinless theory.\medskip

In Sections \ref{porzadeknabiurku}
and \ref{families} we explain that it can be assumed w.l.o.g. that the counterexample  satisfies some additional assumptions.

The additional assumptions from Section  \ref{porzadeknabiurku} concern trivial simplifications of $\cal T_C$ and $\dd_C$.
One of them is that $\dd_C$ is the $\emptyset$.

The assumptions from  Section \ref{families} however can hardly be seen as simplifications and are one of the main ideas of the whole proof. We define 
{\em family patterns} there, and show that it can be assumed w.l.o.g. that $\cal T_C$ respects the family patterns and that this assumption is a useful tool
giving some insight into the structure of Chase. 

Then, in Sections \ref{in} -- \ref{ch} we show that if ${\cal T}_C$ and $\Phi_C$ satisfy the assumptions from 
Sections \ref{porzadeknabiurku}
and \ref{families},
then the triple ${\cal T}_C$, $\emptyset$, $\Phi_C$ cannot be a counterexample. We lack language to discuss it yet, so the general architecture of this part of 
the proof will be  described in Section \ref{in}.

\vspace{-2mm}

\section{Some trivial simplifications}\label{porzadeknabiurku}

Nothing deep is going to happen here. We are just cleaning our desk before the real work starts.
Our feelings will not be hurt if the reader chooses to read only Lemma \ref{emptyde}, first 3 lines of subsection \ref{piectrzy},
first 10 lines of subsection \ref{pieccztery} and the very short subsection \ref{piecpiec}, and skip the rest of this Section.

\subsection{Empty $\dd$}

\lemmaaa{\label{emptyde}
There exists a counterexample ${\cal T}, \emptyset, \Phi$ for FC.
}

\noindent
{\em Proof:}
Suppose the active domain of $\dd_C$ is $\{d_1, d_2, \ldots d_m\}$
Add a new relation symbol $D$ of arity $m$ 
to the signature of ${\cal T}_C $. Let $\cal T$ consist of all the rules of ${\cal T}_C $, of the rule:
$$ \Rightarrow \exists x_1, x_2\ldots x_m \; D(x_1, x_2, \ldots x_m)$$

and of one Datalog rule:

$$D(x_1, x_2, \ldots x_m) \Rightarrow R(x_{i_1},x_{i_2}\ldots x_{i_k})$$ 

for each atom $R(d_{i_1},d_{i_2}\ldots d_{i_k})$ true in $\dd_C$.

Then clearly  ${\cal T}$, $\emptyset$, $\Phi$ is  a counterexample for FC.\eop

From now on we assume, w.l.o.g. that the triple ${\cal T}_C $, $\emptyset$, $\Phi_C$ is a counterexample for FC.

\subsection{Handy lemma}

\newcommand{\mae}{\text{\ae}}

In this and the next Sections we are going to ''normalize'' theory ${\cal T}_C$. This will be done in several steps. 
The general idea of each of those steps will be that the predicates of ${\cal T}_C$ will be ''annotated'', so that the name of predicate 
will carry some additional information. This will lead to a new signature and a new theory, and in each case we will prove a ''simplifying lemma''
saying that the new theory (together with some new query, and with empty database) is still a counterexample for FC.

In this subsection we present a technical lemma which is a workhorse exploited  in the proofs of all the simplifying lemmas in Sections  \ref{porzadeknabiurku}
and \ref{families}.

\definitionnn{\label{wydlubacpredykat} For an atomic formula $Q=R(\bar t)$ over a signature $\Sigma$ we define  $Q_{|\Sigma}$ to be $R$.
}

In other words  $Q_{|\Sigma}$ is the predicate symbol of $Q$.

\definitionnn{\label{datablind}
Let \ae ~be a function from the set of all atoms over some signature $\Sigma_A$ to the 
set of all atoms over signature $\Sigma$. We will say that \ae ~is annotation erasing if:
\begin{itemize}
 \item[(i)] $\forall C,C'~~ C_{|\Sigma}=C'_{|\Sigma} ~\Rightarrow ~ $\mae$(C)_{|\Sigma_A}=\mae(C')_{|\Sigma_A}$
 \item[(ii)]  $\forall P\in \Sigma_A~\forall i~\exists j~\forall C~~~ C_{|\Sigma_A} = P~ \Rightarrow ~  C(j)=\mae(C)(i)$
 \item[(iii)] $\forall P\in \Sigma_A~\forall i~\exists j~\forall C~~~ C_{|\Sigma_A} = P~ \Rightarrow ~  C(i)=\mae(C)(j)$ 
 \item[(iv)] $\mae$ is onto.
\end{itemize}

For an annotation erasing \ae ~and any formula (or any structure) $X$ by $\mae(X)$ we mean the formula (structure) being the result of replacing each 
atom $Q$ in $X$ by $\mae(Q)$. 
}

See that the above definition requires \ae ~to be ''data blind'':
Condition (i) says that the predicate symbol of the atom being the output of \ae ~must only depend on the  predicate symbol of the input.
Conditions (ii) and (iii) say that all \ae ~is allowed to do is to copy data to the new atom, without really reading them, without inventing 
new data and without forgetting anything. It can however change the order of arguments, and possibly create, in the output relation,  many columns 
being a copy of a given column in the input. Notice that the domain of \ae ~is the set of all atoms -- both ground atoms and atoms containing variables.

\observationnn{\label{komutuja}If \ae ~is  annotation erasing and $h$ is a valuation of variables then the equality $\mae\circ h=h\circ\mae$ holds.\eop}

\definitionnn{ \label{f_inverse}
Let \ae ~be annotation erasing. 
\begin{itemize}
 \item[(i)] The preimage of an atom $C$ under \ae  ~is defined as the disjunction  \ae$^{-1}(C)=\bigvee_{\text{\ae}(B)=C}B$.
 \item[(ii)] The  preimage of a CQ $\Phi=\exists\bar x\bigwedge_i~C_{i}(\bar x)$ under $\mae$ is defined as the UCQ $\mae^{-1}(\Phi)=\exists\bar x\bigwedge_i~\mae^{-1}(C_{i}(\bar x))$.
 \item[(iii)] The  preimage of a UCQ $\Phi=\bigvee_i\Phi_i$ under $\mae$ the UCQ is defined as $\mae^{-1}(\Phi)=\bigvee_i~\mae^{-1}(\Phi_{i})$.
\end{itemize}

}

Notice that correctness of the above definition follows from condition (iii) of Definition \ref{datablind} -- since $\mae$ is not allowed 
to forget an argument, the preimage-image of an atom is always finite.

\lemmaaa{\label{composition}
For a conjunctive query $\Phi$ and annotation erasing $\mae$ the query $\mae(\mae^{-1}(\Phi))$ is equivalent to $\Phi$.}

\noindent
{\em Proof:} Because $\mae^{-1}$ is applied to each atom separately, it is enough to show that, for an atom $C$, $\mae(\mae^{-1}(C))$ is equivalent to $C$. 
By definition we have 
$$\mae(\mae^{-1}(C))=\mae(\bigvee_{\mae(B)=C}B)=\bigvee_{\mae(B)=C}\mae(B)=\bigvee_{\mae(B)=C}C= C$$

The first equality is a direct application of Definition \ref{f_inverse}(i).  The second equality is a direct application of Definition \ref{f_inverse}(iii).
 In the last equality we used the fact, that for each atom C there exists at least one atom B such that $\mae(B)=C$. But this follows from Definition \ref{datablind} (iv).

\definitionnn{\label{annotation} For a given theory $\cal T$ over a signature $\Sigma$, and an annotation erasing $\mae$,
a theory  ${\cal T}_A$ over a signature $\Sigma_A$ is called an $\mae$-annotation of  $\cal T$   if 
for each rule $\Phi\Rightarrow Q$ in ${\cal T}$ (resp. $\Phi\Rightarrow \exists z Q$ in $ {\cal T}$) and each conjunction of atoms $\Phi'$ such that $\mae(\Phi')=\Phi$ 
there exists exactly one rule  $\Phi'\Rightarrow Q'$ in ${\cal T}_A$ (resp. $\Phi'\Rightarrow \exists z Q' $ in $ {\cal T}_A$) such that $\mae(Q')=Q$.
}

The sense of the definition is that the new theory contains annotated versions of the rules of the old theory. There is exactly one new rule 
for each possible annotation of atoms in the body of an old rule. 

\lemmaaa{\label{chase_equal}
If ${\cal T}_A$ is an $\mae$-annotation of ${\cal T}$  then $Chase({\cal T},\emptyset)=\mae(Chase({\cal T}_A,\emptyset))$.}

\noindent
{\em Proof:} By induction one can easily show that $Chase^{i}({\cal T},\emptyset)=\mae(Chase^{i}({\cal T}_A,\emptyset))$. 
The induction step follows directly from Definition \ref{annotation}. Notice that for the $\supseteq$ inclusion 
 the phrase ``exactly one'' in Definition \ref{annotation} is crucial.\eop

\lemmaaa{\label{homomorphism}
For an annotation erasing $\mae$ and a CQ $\Phi$, 
if $M\models \Phi$ then $\mae(M)\models \mae(\Phi)$.}

\noindent
{\em Proof:} Let $h$ be a valuation of $Var(\Phi)$ which shows that $M\models \Phi$. In other words, the image of $\Phi$ under $h$  is a substructure of $M$ i.e. $h(\Phi)\subseteq M$.  
Hence, $\mae(h(\Phi))\subseteq \mae(M)$.
By Observation \ref{komutuja} we get $\mae\circ h=h\circ \mae$, so $h(\mae(\Phi)) \subseteq \mae(M)$. Therefore $\mae(M) \models \mae(\Phi)$.\eop

\lemmaaa{[Handy Lemma]\label{handy} If the triple ${\cal T},\emptyset, \Phi$ is a counterexample to FC and ${\cal T}_A$ is an  $\mae$-annotation of ${\cal T}$, then the triple ${\cal T}_A, \emptyset, \mae^{-1}(\Phi)$ is also a counterexample for FC.}

\noindent
{\em Proof:} Suppose $Chase({\cal T}, \emptyset) \not\models \Phi$.
Lemma \ref{chase_equal} states that  $Chase({\cal T},\emptyset)=\mae(Chase({\cal T}_A, \emptyset))$ and Lemma
\ref{composition} states that $\Phi=\mae(\mae^{-1}(\Phi))$,
so by contraposition of Lemma \ref{homomorphism} we get $Chase({\cal
T}_A, \emptyset)\not \models \mae^{-1}(\Phi)$.

Let $M$ be an arbitrary finite model of ${\cal T}_A$. We need to show
that $M\models \mae^{-1}(\Phi)$. Because ${\cal T}_A$ is an annotation
of ${\cal T}$, we have that
 $\mae(M)$ is a model of ${\cal T}$.
Hence, $\mae(M)\models \Phi$ -- this is because $({\cal T},\emptyset,
\Phi)$ is a counterexample to FC, so $\Phi$ must be satisfied in each
finite model of $\cal T$.

Let $\Phi_0=\exists\bar x~\Psi (\bar x)$ be a disjunct of $\Phi$ which is true in $\mae(M)$. 
 There exists $h$  - a valuation of the variables $\bar x$ - such that $h(\Psi(\bar x))\subseteq \mae(M)$.
This inclusion implies that there exists a subset $M_0$ of $M$ (we see $M$ as a set of atoms now) such that $h(\Psi(\bar x))=\mae(M_0)$.

Now we claim that $M_0\models \mae^{-1}(\Phi)$. Of course when we prove this claim then the proof of Handy Lemma will be finished.  
But, by definition of preimage we have that $\mae^{-1}(\Phi_0)$ logically implies $\mae^{-1}(\Phi)$, so it will be  enough to
notice that  $M_0\models \mae^{-1}(\Phi_0)$. Again using definition of preimage (and distributivity) we see  that
 $\mae^{-1}(\Phi_0)$ is a disjunction of all possible CQs $\exists \bar x \Psi_0$ such that $\mae(\Psi_0)=\Psi$. 
And $M_0$ is a homomorphic image of one such $\Psi_0$.\eop


\subsection{Strongly joinless theories}\label{piectrzy}

We will call a joinless theory $\cal T$ {\em strongly joinless} if heads of all the rules of $\cal T$ are joinless, which means that if $T$ is a rule from $\cal T$ then 
each variable occurs in the head of $T $ at most once.

\lemmaaa{\label{veryjoinless}
 There  exists  a  counterexample  ${\cal T}_A$, $\emptyset$, $\Phi_A$ for FC with 
 strongly joinless   ${\cal T}_A$. 
}

%

\noindent
{\em Proof :}

\newcommand{\caltaa}{{\cal T}_{A}}
\newcommand{\sigmaa}{\Sigma_A}
\newcommand{\phiaa}{\Phi_A}

%
%
%
%

\noindent
{\bf Annotations.} For a natural number $k$ by a $k$-annotation we will mean any set of equalities of the form $i=j$ for $1\leq i,j\leq k$ which is closed under logical consequence. 
For $k$-annotations $\alpha_1$ and $\alpha_2$ by $\alpha_1 \wedge \alpha_2$ we will mean the smallest annotation containing  $\alpha_1$ and $\alpha_2$. For 
a $k$-annotation $\alpha $ by index of $\alpha$ we will mean the number of equivalence classes that $\alpha$ 
naturally splits $\{1,2,\ldots k\}$ into. \vspace{1mm}

\noindent
{\bf The signature of $\caltaa $.}
Let $\Sigma$ be the signature of ${\cal T}_C$. The signature $\sigmaa $ of $\caltaa $ will consist of one predicate 
$R_\alpha$ for each $k$-ary predicate $R\in \Sigma$ and for each $k$-annotation $\alpha$. The arity of $R_\alpha$ equals to the index of $\alpha$.
The following notational convention will apply: Atoms of $R_\alpha$ will be written as $R_\alpha(X_1,X_2,\ldots X_k)$ with $X_i$ being equal to $X_j$ whenever $i=j$ is in $\alpha$.
For example if $k=3$ then $R_{2=3}(a,b,b)$ is an atom of the binary relation $R_{2=3}$ while the expression $R_{2=3}(x,x,y)$ is a ({\footnotesize $\heartsuit$}) syntax error .\vspace{1mm}
%
%

\noindent
{\bf Theory $\caltaa$.} Now we are ready to define theory $\caltaa$. For each (joinless) rule of  ${\cal T}_C$: \vspace{1mm}

$T$\hfill $R(x_1,x_2\ldots x_k), P(y_1, y_2\ldots y_m)\Rightarrow \exists z\; Q(v_1,v_2,\ldots v_l)$ \vspace{1mm}

where each of $v_i$ is either $z$ or one of the $x_i$ or one of the $y_i$, and for each $k$-annotation $\alpha$ and each $m$-annotation $\beta$, theory  $\caltaa$
will contain the rule: \vspace{1mm}

${T}_{\alpha,\beta}$\hfill $R_\alpha(X_1,X_2\ldots X_k), P_\beta(Y_1, Y_2\ldots Y_m)\Rightarrow 
\exists z\; Q_\gamma(V_1, V_2,\ldots V_l)$ \vspace{1mm}

\noindent
where: \vspace{1mm}

{\bf --}$X_i=X_j$ if $i=j$ is in $\alpha$ and $Y_i=Y_j$ if $i=j$ is in $\beta$;

{\bf --} $V_i$=$X_j$ if $v_i=x_j$, $V_i$=$Y_j$ if $v_i=y_j$, and $V_i$=$V_j$ if $v_i=v_j$;

{\bf --} $i=j$ is in $\gamma$ if and only if $V_i=V_j$. \vspace{1mm}

Notice that the rule ${T}_{\alpha,\beta}$ is strongly joinless -- arity of $Q_\gamma$ is equal to the index of $\gamma$ and is equal to the number of 
different variables in the atom $Q_\gamma(V_1, V_2,\ldots V_l)$.

To keep the notation as simple as possible we defined ${T}_{\alpha,\beta}$ for a TGD with two atoms in the body. But of course the same must be done for
all rules of ${\cal T}_C$, including Datalog rules.\vspace{1mm}

\noindent
{\bf Annotation erasing.} Now \ae ~is defined as an operation that maps atoms over signature $\sigmaa$  to atoms over $\Sigma$, 
in the most natural way one could imagine -- by erasing the annotation.

It is easy to notice that \ae ~is indeed an annotation erasing, as defined by Definition \ref{datablind}, and that $\caltaa$ satisfies the assumptions of Handy Lemma. 
So, we can use  Handy Lemma to finish the proof of of Lemma \ref{veryjoinless}. \eop

From now on we will assume, w.l.o.g. that the triple ${\cal T}_C, \emptyset, \Phi_C$ is a strongly joinless counterexample for FC.

\subsection{Almost clean theories}\label{pieccztery}

We will call a strongly joinless theory  $\cal T$ {\em almost clean} if
each rule from
$\cal T$ is either a Datalog rule of the form: 
\trefle{rzutowanie}{$Q(\bar x)\Rightarrow Q'(\bar x^i)$, 
where by $ \bar x^i$ we mean  the tuple $ \bar x$ with $i$-th element removed}\medskip

or a TGD of the form:\medskip

\trefle{siedziecporzadnie}{
$Q_0(\bar x_0)\wedge Q_1(\bar x_1)\Rightarrow  \exists y \; Q(y,\bar x_0, \bar x_1)$ 
}\medskip

Condition  \rtrefle{siedziecporzadnie} does not rule out  $Q_1$ to be empty, so in particular a rule of the form $Q_0(\bar x_0)\Rightarrow  \exists y \; Q(y,\bar x_0)$ is also allowed. 
The important part, of both conditions, is that the  variables in the head occur in exactly the same  order as the  variables in the body.

\lemmaaa{\label{veryjoinless}
 There  exists  a  counterexample  ${\cal T}_A$, $\emptyset$, $\Phi_A$ for FC with 
  ${\cal T}_A$  being almost clean. 
}

\noindent
{\em Proof:} It is trivial to see that  ${\cal T}_C, \emptyset, \Phi_C$ can w.l.o.g. be assumed to contain only TGDs with at most two atoms in the body 
(the left hand side) and of Datalog rules that project exactly one element. The slightly more non-trivial part is to show that the ordering condition can also be satisfied.

\noindent
{\bf The signature of $\caltaa $.}
Let $\Sigma$ be the signature of ${\cal T}_C$. The signature $\sigmaa $ of $\caltaa $ will consist of one predicate 
$R_\alpha$ for each $k$-ary predicate $R\in \Sigma$ and for each $k$-permutation $\pi: \{1,\ldots k\}\rightarrow  \{1,\ldots k\}$. The arity of $R_\pi$ is equal to the
arity of $R$.

\noindent
{\bf Annotation erasing}  \ae ~is defined as $\mae(Q_\pi(x_1,x_2,\ldots x_k)= Q(\pi(\bar x))$, where $\pi(\bar x) = (x_{\pi(1)},\ldots x_{\pi(k)})$. Clearly, this 
operation satisfies the requirements
of Definition \ref{datablind}.

\noindent
{\bf Theory $\caltaa$.} For each Datalog rule $T$ of theory ${\cal T}_C$, of the form
 $Q(\bar x)\Rightarrow Q'(\pi_T(\bar x^i))$, with $Q$ of some arity $k$,  and for each $k$-permutation $\pi$
let there  be a rule $T_\pi$ in ${\cal T}_C$, of the form $Q_{\pi}(\bar x)\Rightarrow Q'_{\pi'}(\bar x^i)$, where $\pi'$ is the unique permutation such that
$T$ equals, up to the renaming of variables, to \ae$(T_\pi)$. In similar manner we construct  one rule for each TGD in ${\cal T}_C$ and each possible annotation
of the predicates in the body of this rule. Then use Handy Lemma to finish the proof.\eop

\subsection{Clean theories and clean counterexamples}\label{piecpiec}

An almost clean theory  $\cal T$ will be called {\em clean} if:

\begin{itemize}
\item the signature of $\cal T$ is a union of two disjoint sets: parenthood predicates (or PPs),
occurring
in the heads of rules of the form  \rtrefle{siedziecporzadnie}, and projection predicates,  occurring
in the heads of rules of the form \rtrefle{rzutowanie};

\item
for each projection predicate $Q$ there is a parenthood predicate $Q'$ such that  
$Q(\bar t) \Rightarrow \exists t\; Q'(t, \bar t)$ and $Q'(t,\bar t) \Rightarrow Q(\bar t)$ are
rules of $\cal T$.
\end{itemize}

We will call a UCQ (or a CQ) $\Phi$  {\em clean} if only the parenthood predicates appear in $\Phi$.
A triple  $\cal T$, $\emptyset$, $\Phi$  is {\em clean} if $\cal T$ and $\Phi$ are  clean.
Using Lemma \ref{veryjoinless} it is very easy to show that:

\lemmaaa{There exists a clean counterexample $\cal T$, $\emptyset$, $\Phi$ for FC.\eop
}


\section{On the importance of family values}\label{families}

Let $\cal T$ be a clean theory, as defined in Section \ref{porzadeknabiurku}. From now on 
we will always have $\dd=\emptyset$. Since
the context is clear we will simply write $Chase$ instead of $Chase({\cal T},\emptyset)$. 

In this Section we will  imagine 
$Chase$ as the humankind. Generations after generations of elements are being born
(by the TGDs) and then projected out (by the Datalog rules). And atoms are like families, as you are going to see. 
 Let $l$ be the maximal predicate  arity in the signature of $\cal T$.

\subsection{A fairy tale}

In the next subsection we define {\bf family patterns}. This is a crucial tool in our analysis of the structure of $Chase$, but a complicated one.
So before we present  the technical definitions, the reader is invited to join us for an informal visit to a planet far far away,
where very strict rules apply concerning family dinners. 

First of all,  the participants of a family dinner must always be all the ancestors of some person $A$ (who may be alive or dead at the moment) 
who are currently alive. The word ''ancestor''
is understood in the reflexive sense, which means that $A$ must also participate, if she is still alive. A group of people that are allowed to dine together 
will be called ''family''. 

Due to some  curse  no family on this planet can ever have more than $l$ members. Notice that the families, as we defined them, are not pairwise disjoint. 
Adam, Eve and Cain were  a family. Adam, Eve and Abel were a family, and after Abel's death (but not before) Adam and Eve were still a family. But there was never a family including both Cain and Abel. 

During a dinner all the participants sit behind a long table, always in the same order. When someone is sadly projected out, 
then the surviving family members  shift (so that there is no empty space left), but the order remains the same. 

Two families can sometimes have a baby together. One peculiarity is that all the ancestors of $A$ who were alive when $A$ was born are considered parents of $A$.
When two families have a baby $A$ together then, according to the above rule, they are allowed to dine together. Here is how they are seated during such dinner: $A$ sits first from the left, 
then all the people from the mother family, in the order they used during their dinners, and then all the people from the father family, also in the order they used during their dinners.

Now imagine being the Police who enforce the rules. You come and just see a row of people behind the table, with no apparent structure at all.  This is why
a rule was introduced requiring that each family posts information about their {\em family ordering} on their web page. Family ordering is the binary relation (actually a partial order), 
on elements
$\{ 1,2,\ldots k\}$ (where $k$ is the cardinality of the family) containing all the (descendant, ancestor) pairs. Notice that people are identified here with the places they occupy
when dining. Since $k\leq l$ there are only finitely many possible family orderings. 

Another aspect of the family life that is strictly codified is the way people address their parents (ancestors). When a newborn $A$ dines, for the first time,
with all her parents (i.e. living ancestors) she learns to call the person $B$  sitting on the chair $i$ as simply $i$. Then, as time goes by, some of $A$'s ancestors are projected out,
 new people are being born, but $A$ and $B$ may still dine together, in different configurations. And  $A$ will always address $B$ as $i$. 

The function that maps each (descendant, ancestor) pair (C,D) of a family to a number (not greater than $l$) used by $C$ to address $D$ is also posted on the family web page. 
Together with the family ordering they form the {\bf family pattern}. Notice that there are only finitely many possible family patterns.\vspace{1mm}

{\bf Remark about incest}. 
There is nothing in the rules of the planet that would forbid non-disjoint families to have a baby together. Actually there is nothing in the rules that would require that 
mother family and father family are different (which some humans may see as strange). And when two non-disjoint families have a baby then 
there is  a person who plays more than one role -- he is
 a member of the mother family and of the father family at the same time. Such person has two (or more) chairs behind the family table, and the way he is addressed by his descendants 
depends on the chair he currently sits on. Notice that  the family ordering is defined as a partial order of chairs rather than people and  it is blind to the fact that two chairs are occupied 
by the same person and thus it is always a tree-like ordering -- each two
descendants of a given element are always comparable.\vspace{1mm}

{\bf Back to the example}. As we said, Cain, Eve and Adam were a family. The family pattern was $F,\delta$ where the 
ordering $F$ consisted of two pairs: $ 2<_F 1$ and $ 3<_F 1$ and $\delta$ was defined as $\delta(1,2)=2$ and $\delta(1,3)=3$.

Also  Awan, Eve and Adam were a family. And the family pattern was the same  $F,\delta$ as before.

Then Awan and Cain had a child together, named Enoch. When Enoch was born Adam and Eve were still alive, so the five people were one family.
But there were seven chairs behind the  table they needed to dine together. First chair for Enoch, 2nd for Awan, 3rd for Eve, 4th for Adam,
5th for Cain, 6th again for Eve and 7th again for Adam. The new family pattern was $G, \gamma$ where $G$ consisted of the pairs
$i<_G 1$, for each $2 \leq i\leq 7$ and of the pairs $3<_G 2$, $4<_G 2$, $6<_G 5$ and $7<_G 5$. What concerns $\gamma$, we had
$\gamma(1,i)=i$, for each $2 \leq i\leq 7$, and -- for example -- $\gamma(5,7)=\delta(1,3)=3$, as the way Cain was calling  Adam did not change after Enoch was born.

\subsection{Family patterns and how they change over time}

Let us now formalize our fairy tale:

\definitionnn{
By a ($k$-ary) family ordering we mean any  tree-like partial order, whose set of vertices is 
$\{ 1,2,\ldots k\}$ where $k\leq l$. By a tree-like partial order we mean that each two elements 
greater than any given one
 are comparable.
If a family ordering is a tree then 1 is the root (the greatest element) of this tree.
}

If a family ordering is a tree, the root of the tree is 
the youngest family 
member\footnote{
Mnemonic hint: the one is smaller whose 
date of birth is a smaller number.
}.


But -- as we explained above -- the family ordering alone is not everything we want to know about a family. 
Alice dining only with her granny form the same ordering as Alice dining with her mother,
but they do not form the same family pattern:

\definitionnn{
A ($k$-ary) family pattern is a pair $F,\delta$, where $F$ is a ($k$-ary) family ordering and $\delta $ is a 
function assigning a number, from the set $\{1,2,\ldots l\}$, to each pair $j,i$ of elements of $F$ such that
$i<_F j$, where $<_F $ is the ordering relation on  $F$ ($i$ is an ancestor of $j$).  
}

Clearly, once the maximal arity $l$ is fixed, the set of all possible  family patterns is finite.

Now imagine there had been  a family of $k$ people with the family pattern $F,\delta$. 
But then, at some point of time, the person who sat on chair $i$ was projected out. 
The surviving family members still dine together, and their new family 
pattern is of course a function of $F,\delta$ and of $i$. Call the new pattern\footnote{If you are not happy with this definition, then
treat Observation \ref{project} as a definition. The same applies to Observation \ref{baby}.}  project$_i(F,\delta)$
We will never really need to compute project$_i(F,\delta)$, but maybe it is helpful to see that it is indeed possible:

\begin{observ}\label{project}
Suppose project$_i(F,\delta) = G,\gamma$.
For a natural number $1\leq j\leq k$ let $g(j)=j$ if $j<i$ and  $g(j)=j-1$ otherwise. Then $j<_F j'$ if and only if $ g(j)<_G g(j')$ and, whenever  
$j<_F j'$ then $\delta(j,j') = \gamma(g(j),g(j'))$
\end{observ} 

In a similar manner we can imagine two families, one consisting of $k$ people, with the family pattern  $F, \delta$, and another one 
with  $k'$ people, and with the family pattern  $F', \delta'$, having a baby together. Then, together with the baby, they form a new 
family, of $1+k+k'$ people, and the family pattern of the new family is a function of $F, \delta$ and $F', \delta'$.  Call the new pattern 
baby$(F, \delta; F', \delta')$. Again, this is not really needed but we can compute baby$(F, \delta; F', \delta')$:

\begin{observ}\label{baby}
Suppose baby$(F, \delta; F', \delta')= G,\gamma$. Then:

\begin{itemize}
\item[(a)]
$i<_G j \;\Leftrightarrow \;
(j=1 \wedge i>1)\;\vee  $\\ \hspace*{16mm}
$(i-1<_F j-1 \;\;\; \wedge \;\;\; 1< i,j\leq k+1)\;\vee $\\ \hspace*{16mm}
$(i-k-1<_{F'} j-k-1 \;\;\; \wedge \;\;\; k+1<i,j\leq k+k'+1)  $

\item[(b)]
If  $j=1$ and $1<i\leq k+k'+1$ then $\gamma(j,i)=i$.\\
If $1<j,i\leq k+1$ then $\gamma(i,j)=\delta(i-1,j-1)$.\\ 
If $k+1<j,i\leq k+k'+1$ then $\gamma(i,j)=\delta'(i-k-1,j-k-1)$. 

\end{itemize}
\end{observ} 

Condition (a) says that the birth of the new child does not change the ancestor relation in the family,
except from the fact that
each of the members of the two families is now also this child's ancestor. The meaning of condition (b) is that 
the newborn child  learns how  to address his ancestors: 
it addresses them  by their positions at the family table, as it sees 
it at the moment of its  birth. The child's birth does not change the way
his ancestors are addressing each other.


\subsection{Back to the Chase}
%

\definitionnn{\label{chinskieznaczki}
A clean theory {\em $\cal T$ respects family patterns} if:

\begin{enumerate}
\item Each relation $Q$ of arity $k$ in the signature of $\cal T$ contains, as a part
of its name (as a subscript) a $k$-ary family pattern.

\item If $R_{F,\delta}(\bar x) \Rightarrow P_{G,\gamma}(\bar x^i)$ is a Datalog rule of $\cal T$ then $G,\gamma = $project$_i(F,\delta)$ (the meaning of $\bar x^i$ is as defined in subsection \ref{pieccztery}).

\item If 
$R_{F,\delta}(\bar x) \wedge R'_{F',\delta'}(\bar x') \Rightarrow \exists y\; P_{G,\gamma}(y,\bar x, \bar x')$ is a TGD of $\cal T$ then we have $G,\gamma$=baby$(F, \delta; F', \delta')$

\end{enumerate}  
}

\lemmaaa{
There exists a clean counterexample  
${\cal T}_A$, $\emptyset$, $\Phi_A$, with 
 ${\cal T}_A$ respecting family patterns.}

\noindent
 {\em Proof:} Let ${\cal T}_C$, $\emptyset$, $\Phi_C$ be any clean counterexample, over some signature $\Sigma$.
Let $\Sigma_A$ consist of one arity $k$ predicate $Q_{F,\delta}$ for each arity $k$ predicate $Q$ in $\Sigma$ and each 
$k$-ary family pattern $F,\delta$. 

Now for each Datalog rule 
$R(\bar x) \Rightarrow P(\bar x^i)$ in ${\cal T}_C$ and for each family pattern $F,\delta$ of arity equal to the arity of $R$, 
let 
$R_{F,\delta}(\bar x) \Rightarrow P_{G,\gamma}(\bar x^i)$ be a rule in ${\cal T}_C$, where $G,\gamma = $project$_i(F,\delta)$.

Similarly, for each TGD 
$R(\bar x) \wedge R'(\bar x') \Rightarrow \exists y\; P(y, \bar x, \bar x')$
in ${\cal T}_C$ and for each pair of family patterns $F,\delta$, $F',\delta'$, of arities equal to the arities of $R$, $R'$ respectively,
let 
$R_{F,\delta}(\bar x) \wedge R'_{F',\delta'}(\bar x') \Rightarrow \exists y\; P_{G,\gamma}(y, \bar x, \bar x')$ 
be a rule in ${\cal T}_C$, where $G,\gamma=$baby$(F, \delta; F', \delta')$. Define the function \ae ~as -- literally -- removing the annotations. Use Handy Lemma to finish the proof.\eop

\begin{center}
{\bf \small From now on we assume that\\ $\cal T$ is a fixed clean theory which respects family patterns. } 
\end{center}

Before we end this Section let us
study some properties of $Chase({\cal T},\emptyset)$.
The following Lemma is an obvious consequence of the assumption that $\cal T$ is clean and of freeness of the Chase:

\lemmaaa{
For each element $a$ of $Chase$ there exists exactly one parenthood predicate atom 
$A= PP(a,\bar a)$ such that $Chase \models A$. It will be called {\em the parenthood atom}
of $a$, and the elements of $\bar a$ will be called parents of $a$. 
}

Notice that we use the word ''parents'' (here and always in the future) to denote all the ancestors of $a$ who were present when $a$ was born.  So it 
it is perfectly normal in our scenario that $a$ and $b$ are parents of $c$ while $a$ is a parent of $b$.

\definitionnn{\label{zero}
For two elements $a,b$ of $Chase$ we will say that $a$ and $b$ are 0-equivalent (denoted $a \equiv_0 b$)
if the parenthood atoms of $a$ and $b$ are atoms of the same predicate. 

Suppose $a \equiv_0 b$, and $A$ and $B$ are parenthood atoms of $a$ and $b$ (resp.). Then, for each $i$,
the pair of elements $A(i)$ and $B(i)$ will be called {\em respective parents} of the pair of elements $a$ and $b$. 
For tuples $a_1$,$a_2,\ldots$ $a_s$ and $b_1$,$b_2,\ldots$ $b_s$ by $a_1$,$a_2,\ldots$ $a_s\equiv_0$ $b_1$,$b_2,\ldots$ $b_s$
we mean that $a_i\equiv_0 b_i$ for all $1\leq i\leq s$. 
}

Since the family pattern is part of the name of the predicate, when we say ''the same predicate'' in Definition \ref{zero} we of course mean 
that the family patterns are also equal.

The next lemma  says, using our
running metaphor, that the person an element $a$ of $Chase$ calls its granny does not change during its lifetime.
Moreover, the way  $a$'s father calls $a$'s granny also remains unchanged:

%
%

\lemmaaa{\label{staloscnazw}
Suppose $Chase \models B,C$, for $B=Q_{F,\delta}(\bar b)$ and $C=PP_{G, \gamma}(a,\bar a)$.
Suppose also that $a=B(i)$ and  $ j,j'<_F i$. Then:

\begin{enumerate}
\item
 $B(j)$ is  a parent of $a$;
\item
$B(j) = C(\delta(i,j))$;
\item
$j <_F j'$ if and only if $\delta(i,j) <_G \delta(i,j')$;
\item 
if $j <_F j'$ then $\delta(j',j)= \gamma(\delta(i,j'),\delta(i,j))$.
\end{enumerate}
}

The proof of the lemma is easy induction on the structure of $Chase$, and we leave it for the reader as an exercise. 
Actually, the only possibly non-trivial part of this exercise is  to remember what the notations mean. So let us come to your help. 
The assumption that $a=B(i)$ means that $a$ is somewhere (position $i$'th) in atom $B$. The assumption that  $ j <_F i$ means that in family $B$
the element in position $j$, call it $c$,  is (according to the family pattern of this family) an ancestor of $a$. Now trace the history (or ''derivation in $Chase$'') of the family 
(or ''of atom'') $B$
back to $a$'s birth, and notice that each step of the derivation preserves the properties claimed by the Lemma. The last is because we assume that $\cal T$ respects the family patterns.

Now we have something slightly more complicated. The following  lemma, which will be critically important 
in Section \ref{ch}, is where the power of family patterns is seen:

\definitionnn{\label{py}
For a family ordering $F$ and a set $\cal I$ of positions  in $F$ we define
the set $PY({\cal I})$ of positions in $F$ as $\bigcap_{i\in {\cal I}} \{j\in F:\neg (j\leq_F i)\}$.
}

$PY({\cal I})$ (which reads ''possibly younger'')  is exactly the set of family members who
 potentially can be younger than each of the elements of $\cal I$. Of course the set $PY$ depends on the ordering $F$,
but we do not make it explicit in the notation as the context is always clear.

\lemmaaa{[About the Future]\label{future1}

Let $Chase\models A$ for some
$A=PP_{F,\delta}(a,\bar a)$. Suppose
${\cal I}=\{i_1,i_2,\ldots i_s\}$ is a set of pairwise $<_F $-incomparable  positions in $F$ and let 
$b_1,b_2,\ldots b_s$ be equal to  $A(i_1),A(i_2),\ldots A(i_s)$ respectively. 
Suppose  $d_1,d_2,\ldots d_s$ is another tuple 
of elements of $Chase$ such that $b_1,b_2,\ldots b_s  \equiv_0 d_1,d_2,\ldots d_s$. Then there
exists an atom $C= PP_{F,\delta}(c,\bar c)$, such that:

\begin{enumerate}

\item[(i)]
 $Chase\models C$;

\item[(ii)]
$d_1,d_2,\ldots d_s$ equal $C(i_1),C(i_2),\ldots C(i_s)$ respectively;

\item[(iii)]
if $j\in PY({\cal I})$ then $A(j)\equiv_0 C(j)$;
\end{enumerate}
}

Lemma \ref{future1} says that the	 potential of forming atoms in  $Chase$ only depends on the 
$\equiv_0$ equivalence class
of elements (and tuples of independent elements), not on the elements themselves. 
If  $b_1,b_2,\ldots b_s$ and $d_1,d_2,\ldots d_s$ are 0-equivalent tuples of elements and $b_1,b_2,\ldots b_s$
appear in some atom $A$ in $Chase$ (at independent positions)  then there exists an atom $C$, somewhere in $Chase$,
which  not only has $d_1,d_2,\ldots d_s$ in the same positions, but also is 
as similar to $A$ as one could dream of: everything that happens in the future of some $b_i$ in $A$ is 
0-equivalent to the respective future of the respective $d_i$ in $C$. 

Before we prove Lemma \ref{future1}, as one more exercise let us show that it follows easily from 
Lemma \ref{staloscnazw} that if $j\not\in PY({\cal I})$
then the elements  $A(j)$ and $C(j)$ are respective parents of some $b_k$ and $d_k$:

\lemmaaa{\label{future2}
If $i_k\in {\cal I}$ and $j<_F i_k $ then $ A(j) $ and $C(j)$ are respective parents of  
$b_k$ and $d_k$ (where the notations are like in Lemma \ref{future1}). \eop
}

\noindent
{\em Proof of Lemma \ref{future2}.}  Let $A'$ be the parenthood atom of $b_k$ and let $C'$ be the parenthood atom of $d_k$. 
Of course $A'$ and $C'$ are atoms of the same predicate, as we assumed that $b_k\equiv_0 d_k$. 
Then, by Lemma \ref{staloscnazw}.2. we have $A(j)=A'(\delta(i_k,j))$ and  $C(j)=C'(\delta(i_k,j))$, where $\delta$ is as in Lemma \ref{future1}.  \eop\vspace{1mm}

Remember that the fact that $b \equiv_0 d$
does not imply that the respective parents of $b$ and $d$ are 0-equivalent.\vspace{1mm}


\noindent
{\em Proof of Lemma \ref{future1}.} The intuition is that we will trace the genealogy of atom $A$, as deep to the past as we see families 
containing one of the $b_i$. If we go many enough generations back in time we will see, for each $i$, the family in which  $b_i$ was born. Since we assume that 
$b_1,b_2,\ldots b_s  \equiv_0 d_1,d_2,\ldots d_s$ we can find, for each $i$, another atom, somewhere in $Chase$, of the same predicate (including family pattern),
  which gave birth to $d_i$.
Now we can tell the families where $d_i$ were born: ''mimic the behavior of the parenthood atoms of $b_i$''. And they can do it, because the rules are joinless,
which implies that all atoms of the same predicate are equally able to participate in derivations.  
 
\newcommand{ \sv }{$s\hspace{-0.7mm}v\;$}

To be more precise, we  consider (a fragment of) the derivation tree of the atom $A$ in $Chase$, which we will call 
 $\cal D$. Verticies of $\cal D$ will be atoms of $Chase$, with $A$ being the root.  $\cal D$ is defined by induction, together with an equivalence 
 relation  \sv  (as ''same variable'') on the set of all positions in the atoms of $\cal D$, and with the set of painted positions:

\begin{itemize}
\item Atom $A$ is the root of $\cal D$ (and thus an inner node of $\cal D$). 
Positions $i_1,i_2,\ldots i_s$ in $A$ are {\em painted}.

\item 
Suppose an atom 
$B = Q_{ G, \gamma  }(e,\bar e)$ is a node of $\cal D$ 
with  some non-root position painted\footnote{Recall that the root of a parenthood atom is its position 1 -- the root of the family ordering,
which is a tree. An atom which is not a PP-atom may or may not not be a tree and thus it is possible for it to contain only non-root positions.}.  
 Suppose $B' = Q'_{ G', \gamma'  }(\bar {e_1})$ and
$B'' = Q''_{ G'', \gamma''  }(\bar {e_2})$ are such two atoms, 
true in $Chase$, that $B$ was derived in $Chase$, from $B'$ and $B''$, by a single use of the rule: $X' \wedge X'' \Rightarrow \exists x\; X$, where  
$X'= Q'_{ G', \gamma'  }(\bar {x_1})$,  $X'' = Q''_{ G'', \gamma''  }(\bar {x_2})$ and
$X= Q_{ G, \gamma  }(x, \bar x) $.
Then $B'$ and $B''$ are nodes of $\cal D$, and children of $B$. 

If $X(i)=X'(j)$ (or $X''(j)$), which means that the variables on position $i$ in $X$ and on position $j$ in $X'$ (or $X''$) are equal,   
then the pair of positions $i$ in $B$ and $j$ in $B'$ (or $B''$) is added to the relation  \sv  (and  \sv  is always extended to be an equivalence). A position in $B'$ or $B''$ is painted if it is  \sv  with some previously painted position.

The case when $B$ was derived by a projection rule $X'\Rightarrow X$ is handled analogously\footnote{Notice that $B$ while for $B$ being PP-atoms we can always identify
unique pair $B'$, $B''$ in $Chase$ that led to $B$ in one derivation step, this is not always the case if $B$ is a result of a projection. In such case we take $B'$ to be 
any atom of $Chase$ which led to creation of $B$.}.

\item A node of $\cal D$ with no painted positions is a leaf, called an unpainted leaf. A node
which is a PP atom, and whose only painted position is its root is a leaf of $\cal D$, called a painted leaf. All other nodes of $\cal D$ are
inner nodes.
\end{itemize}

The idea here is that we trace  the derivation of $A$ back to the parenthood atoms of the elements $b_i$. The way
we formulated it was a bit complicated, but we could not simply write ''an atom is a leaf of $\cal D$ if it does not contain any of $b_1,b_2,\ldots b_s$''.  This was due to the fact, that  $b$'s can occur in the derivation not only in meaningful positions -- the positions that lead to $i$'s in $A$, but also in non-meaningful ones, not connected, by the rules of $\cal T$, to any of the $i$'s in $A$.

Now, once we have $\cal D$, we construct  another derivation ${\cal D}'$, with the underlying tree 
isomorphic to $\cal D$, defined as follows:

\begin{itemize}
\item If $B$ is an unpainted leaf of ${\cal D}$ then $h(B)=B$ is  the respective leaf of ${\cal D}'$.

\item If $B$ is a painted leaf of ${\cal D}$, which means that $B$ is the parenthood atom of some $b_i$, and if $E$ is 
the parenthood atom of  $d_i$ then $h(B)=E$ is the respective leaf of ${\cal D}'$ (see Observations \ref{propagacja1}--\ref{propagacja3} if you feel an argument is needed here).

\item If $B$ is an  inner node of $\cal D$, being a result of applying some rule $T$ from $\cal T$ to
atoms $B'$ and $B''$ (or just to $B'$, if $T$ was a projection) and if we already know  $h(B')$ and  $h(B'')$  then let $h(B)$
be  the result of applying the rule $T$ to the atoms $h(B')$ and $h(B'')$. 
\end{itemize}

Clearly, ${\cal D}'$ is also a part of $Chase$ and $h(B)$ is always an atom of the same predicate as $B$.
Notice however that if $\cal T$ was not joinless, the last step of the construction would not always  be possible in $Chase$.
 
Now, the atom $h(A)$ in the root of ${\cal D}'$ is going to be the $C$ from the Lemma. What remains to be proved is that 
it indeed satisfies conditions (ii) and (iii) from the Lemma.

It  easily follows from the construction that:

\begin{observ}\label{propagacja1} If $B$, $B'$ are atoms of $\cal D$ and the pair of positions $i$ in $B$ and $j$ in $B'$ is in  \sv  then $B(i)=B'(j)$.
\end{observ}

Notice also that, since $\cal T$ is joinless, which means that a variable in the head of a rule occurs in at most one atom in the body of this rule, we have:

\begin{observ}\label{propagacja2} For an atom $B$ in $\cal D$ and position $i$ in $B$, the set of nodes of $\cal D$ which contain some position being  \sv  to position $i$ in $B$ is a 
 directed path in $\cal D$. 
\end{observ} 

Since, for a $B$ in $\cal D$ we add children of $B$ to $\cal D$ as long as $B$ has some non-root position painted, it follows from the construction that:

\begin{observ}\label{propagacja3} For each position $i_j\in \cal I $ there is exactly one leaf $B$ of $\cal D$ such that root of $B$ and position $i_j$ in atom $A$ are  \sv .
\end{observ}

{\bf Condition (ii).} First of all notice that, as ${\cal D}'$ is isomorphic to $\cal D$, the relation  \sv  can be in a natural way seen as a relation on positions in 
${\cal D}'$ (positions $i$ in $h(B)$ and $j$ in $h(B')$ are  \sv  iff positions $i$ in $B$ and $j$ in $B'$ are), and that Observations \ref{propagacja1}--
 \ref{propagacja3}  still hold true (with $A$ replaced by $C$ in Observation \ref{propagacja3}). 

For $i_j\in \cal I$ consider the leaf $B$ of $\cal D$ such that root of $B$ and position $i_j$ in atom $A$ are  \sv . By Observation \ref{propagacja1} we have
$B(1)= A(i_j)$. Then, by construction of ${\cal D}'$ we have  $(h(B))(1)=d_j$, and, since Observation \ref{propagacja1} remains true in ${\cal D}'$, we have $C(i_j)=d_j$, as needed.

{\bf Condition (iii).} Let $j$ be a position in  $PY({\cal I})$ in $A$. Observation \ref{propagacja2} says that  atoms of $\cal D$ which contain some 
position being  \sv  to position $j$ in $A$ form a 
 directed path in $\cal D$. $A$ is one end of this path. Let $B$ be the other end. There are two possibilities:
either $B$ is unpainted leaf of $\cal D$ or it is an inner node. 

If $B$ is an inner  node, then $B$ is the parenthood atom of $A(j)$. Then $h(B)$ is also the parenthood atom of some element and $B(1)\equiv_0 (h(B))(1)$.
But we know that position 1 in $h(B)$ and position $j$ in $C$ are  \sv , so we have $(h(B))(1)=C(j)$, what needed to be proved. 

If $B$ is unpainted leaf of $\cal D$ then let $i$ be the position in $B$ 
which is  \sv  to position $j$ in $A$. Using the definition of $h(B)$ for the case of unpainted leaves we have $A(j)=B(i)=(h(B))(i)=C(j)$. 
\eop

\section{General scheme of the proof. The first little trick}\label{in}

In the following Sections \ref{in}-- \ref{ch} we show 
that a clean  triple $\cal T$, $\emptyset$, $\Psi$,
where $\cal T$ respects the family patterns,
is never a counterexample for FC.

We will  construct, for our theory $\cal T$,
an infinite sequence of finite structures $\{M_n\}_{n\in \mathbb N}$, which will ''converge'' to $Chase$.
The following property will be satisfied:

\propertyyy{
\label{prop}
\begin{enumerate}
\item[(i)] $M_n \models \cal T $ for each $n \in \mathbb N$.
\item[(ii)]
For each UCQ $\Psi$ and each $n\in \mathbb N$ if $M_n \not\models \Psi$ then  $M_{n+1} \not\models \Psi$.
\end{enumerate}
}

\noindent
Assume -- till the end of this section --  that a sequence  $\{M_n\}_{n\in \mathbb N}$, satisfying Property \ref{prop} (i), (ii) is constructed. Then:

\definitionnn{
A formula $\Phi$ will be called M-true if $M_n\models \Phi$ for each  $n\in \mathbb N $.
}

\lemmaaa{[First Little Trick]\label{flt}
If $\Phi$ is an M-true UCQ
then there exists a disjunct  of $\Phi$ which is M-true.
}

\noindent
{\em Proof:}~ By Property \ref{prop} (ii)  all  queries true in $M_{n+1}$ are also true in $M_n$.
Since $\Phi$ is true in each $M_n$, some disjunct from $\Phi$ must be  true infinitely often,
 and therefore in each $M_n$.\eop\vspace{1mm}

{\bf The rest of the paper is organized as follows:}
In Section \ref{mn} the sequence $M_n$ is defined. 
In Section \ref{secondlt} we present our Second Little Trick, which not only is the main engine of the 
 proof of The Normal Form Lemma but also the main technical idea of the whole paper.

In the very short Section \ref{cyclic} a trivial case of cycled queries (whatever it means) is considered.
 In Section \ref{nfd} we  finally define a normal form of a conjunctive query and explain the main idea of the proof of: 

\lemmaaa{[The Normal Form Lemma]\label{nfl}
For each clean M-true  CQ $\phi$  there exist a clean CQ $\beta$ in the normal form such that:
 
$(*)$ $\beta$ is M-true   and 

$(**)$ $Chase \models (\beta \Rightarrow \phi)$.
}

In Sections \ref{nudna01} and \ref{nudastraszna} we continue the proof of the Normal Form Lemma.
Finally, in Section \ref{ch} we  prove:

\lemmaaa{[The Lifting Lemma]\label{chl}
If a clean CQ $\beta$ is in the normal form and  $M_0 \models \beta$ then $Chase \models \beta$. 
}

Assuming existence of a sequence  $\{M_n\}_{n\in \mathbb N}$, satisfying Property \ref{prop}, and assuming
 Lemmas \ref{nfl} and  \ref{chl}  we can now present:\vspace{2mm}

 \begin{center}{\bf \small The main body of 
the proof of Theorem \ref{maintheorem}:}\end{center}\vspace{-1mm}
Suppose a clean  triple $\cal T$, $\emptyset$, $\Psi$,
where $\cal T$ respects the family patterns,
is  a counterexample for FC. This means there is no finite model satisfying $\cal T$ and not satisfying $\Psi$,
so in particular $\Psi$ is M-true. Let $\phi$ be an M-true disjunct of $\Psi$ (which must exist due to the First Little Trick).
Since  $\cal T$, $\emptyset$, $\Psi$ is a counterexample for FC we know that $Chase({\cal T},\emptyset) = Chase\not\models \Psi$,
so in particular $Chase \not\models \phi$. Let $\beta$ be the normal form of $\phi$ as described in
the Normal Form Lemma.  We know from (*) that $\beta$ is M-true, so in particular  $M_0 \models \beta$.
The  Lifting Lemma tells us that  $Chase \models \beta$. But, by  (**), we have that 
 $Chase \models (\beta \Rightarrow \phi)$, so also  $Chase \models \phi$. Contradiction.\eop\vspace{1.5mm}

The proof above was a high-level one. We neither  bothered to know what the structures $M_n$ are, nor 
what the normal form could actually be. It was enough for us to know
that they are tailored for Lemmas \ref{nfl} and  \ref{chl} to be true.
{\bf The real work begins now.}\vspace{1.5mm}

\section{Aside: a philosophical remark}

We feel we need to address an issue  raised  by one of the reviewers of our LICS submission,
and  explain the connections  between our structures $M_n$ and the
finite structures
from \citep{R06} and \citep{BGO10}.

The general  idea both here and in \citep{R06} and \citep{BGO10} is that a finite structure
constructed by identification of terms which have the same top $n$
levels, for some natural $n$,
can be used to approximate the Herbrand universe with respect to the
properties of elements which only depend on the recent history of
those elements.

This general idea is natural,
by no means new, and  it was reinvented many times by many authors. We
know it from  \citep{M95} (and  its journal version \citep{MP03}), but it is
already present, in some sense, in \citep{JK84}.

The devil is not in the general idea here, but in the details. The
procedure  in \citep{R06} and \citep{BGO10} is the following:

\begin{itemize}
\item start from some database instance $\dd$;
\item give a name to each Skolem function resulting from Skolemization
of the TGDs in $\cal T$;
\item fix $n$ and a constant  $c_n$ to denote the branch stubs;
\item consider the (finite)  universe $U_n$ of the terms of depth up to
$n$ over the defined signature (the constants are $c_n$ and the
constants from $\dd$);
\item define ${\cal T}'$ by replacing each TGD  in $\cal T$ by a
PROLOG rule in the natural way;
run the program ${\cal T}'$ on $U_n$ to get a model  $\bar{U_n}$ of $\cal T$.
\end{itemize}

In this way, the resulting structure is always a model of $\cal T$.
But the cost is that  the $\bar{U_n}$ are very complicated and hard to
analyze. In particular it is not even clear  which atoms  are true in
$\bar{U_n}$, so it is very hard to lift  a valuation satisfying a
query in $\bar{U_n}$ to $Chase(\dd,{\cal T})$.
This is -- if we understand it correctly -- the main source of
complications in \citep{R06} and \citep{BGO10}.

Our way is very much different. We:

\begin{itemize}
\item {\bf first} run $\cal T$ on $\dd$ to get  $Chase(\dd,{\cal T})$;
\item {\bf then}  construct, for given $n$, the structure $M_n$
identifying elements of  $Chase(\dd,{\cal T})$
        which have the same history, up to level $n$.
\end{itemize}

In consequence, not for each $\cal T$ we can be sure that our
structure  $M_n$  is a model of $\cal T$.
But there are two important properties that we get for free, which are
not shared by the structures $\bar{U_n}$ above:

\begin{itemize}
\item $M_n$ is a homomorphic image of $M_{n+1}$. This is a crucial
property for the normalization step
that we call {\em second little trick}.
\item The only way of being an atom in $M_n$ is to be an atom in
$Chase(\dd,{\cal T})$ before. This makes lifting easy.
\end{itemize}

The two above properties make us think  that the structures $M_n$ do
not just {\em approximate} the Chase. They {\em converge} to it.


\section{The canonical models $M_n$ }\label{mn}

Proving that a theory is FC is about building finite models. And finally, in this section we build them. Actually we define  an
infinite sequence of finite models $M_n$, which will  ``converge'' to $Chase$. 

\definitionnn{\label{struktura_chase}
 By  $1$-history of an element $a\in Chase$  (denoted as $H^1(a)$ ) we mean the set consisting  all the parents of $a$.
By the $n+1$-history of $a$ we mean the set $ H^{n+1}(a)=  \bigcup_{b \in H^1(a)} H^n(b) \cup \{b\}$.
 }

Consider now an infinite well-ordered set of colors.
For each natural number $k$ we need to define the $k$-coloring of $Chase$:

\definitionnn{\label{coloring}
 The $k$-coloring is the
coloring of elements of $Chase$, such that each element of $Chase$ has
 the smallest color  not used in its $k$-history. 
}


\definitionnn{\label{rownowaznezero}
For two elements $a$, $b$ of $Chase$  and for $n\in\mathbb{N}$ by $a\simeq_0^n b$ we  mean that
$a\equiv_0 b$ and $a$ and $b$ have the same $n$-color. By $a\simeq_{k+1}^n b$ we mean that
$a\simeq_0^n b$ and that $a'\simeq_{k}^n b'$ for each pair $a',b'$ of respective parents of $a$, $b$.}

To see what  $a\simeq_{n}^n b$ means imagine that 
each element of $Chase$ keeps the record of its family history.
It knows its $n$-color and the name of the predicate it was born with, the $n$-colors of its parents and the 
names of the predicates its parents were born with\footnote{Remember that $a \equiv_0 b $ means that parenthood atoms of $a$ and $b$ are atoms of the same predicate.}.  And so on,
$n$ generations back. Equivalence $\simeq_{n}^n $ identifies elements of $Chase$ if and  only if the records they keep are equal. 

\definitionnn{\label{rownowazne}
For a natural $n\geq 1$ we define two elements  $a,b \in  Chase$ to be $n$-equivalent 
 (denoted as $a \equiv_{n} b$)
if $a\simeq_{k}^k b$ for each $k\leq n$ and if $a'\equiv_{k}  b'$ for each pair $a',b'$ of respective parents of $a$, $b$ and each $k<n$. 
} 


The reader should not feel too much confused by the colors here. They will
only be needed to deal with one trivial case, in Section \ref{cyclic}. Everywhere else
all that needs to be remembered is:

\observationnn{\label{pamietac}The relation $\equiv_n$ is an equivalence relation of finite index. If  $a \equiv_{n+1} b$ then:
\begin{itemize}
\item parenthood atoms of $a$ and $b$ are atoms of the same predicate;

\item $a \equiv_{n} b$;

\item whenever $a'$ and $b'$ are 
 respective parents of $a$  and $b$ then $a \equiv_{n} b$;

\end{itemize}
}

Proof of the Observation is by a straightforward application of Definition \ref{rownowaznezero} and Definition \ref{rownowazne}.

%
%
%

Now the next definition hardly comes as a surprise:

\definitionnn{\label{mstruktury}
Let $M_n$ be the relational structure whose set of elements is $Chase/\hspace{-2mm}\equiv_n$, 
and such that $M_n\models R([a_1],\ldots,[a_n])$ if and only if there are $b_1,\ldots,b_n\in Chase$ such that 
$b_1 \in [a_1], \ldots, b_n \in [a_n]$ and $Chase \models R(b_1,\ldots, b_n)$.
 
}

\noindent

In other words, the relations in $M_n$ are defined in the natural way,  as  minimal with respect to inclusion 
 relations such that the quotient mapping $q_n: Chase\longrightarrow M_n$ is a homomorphism.
If you find  Definition \ref{mstruktury}   complicated, please skip 
Lemma \ref{samodelami} and go first to 
Definition \ref{eva1} and Lemma \ref{eva11} -- we hope they
will shed some light.

Since being $(n+1)$-equivalent implies being $n$-equivalent (Observation \ref{pamietac}) the structure $M_n$ is, for each natural $n$, 
a homomorphic image of $M_{n+1}$, and this implies that
 the sequence of structures $\{M_n\}_{n\in \mathbb N}$  satisfies Property \ref{prop}~(ii). It is also  easy to see that it
 satisfies Property \ref{prop}~(i):


\lemmaaa{\label{samodelami}
 $M_n \models \cal T $ for each $n \in \mathbb N$.
}
\noindent
{\em Proof:} To keep notations as light as possible imagine a rule  $T$ from ${\cal T}$ 
of the form $P(x_1, x_2)\;\wedge \;Q(y_1,y_2) \Rightarrow \exists z~ R(z,x_1,x_2,y_1,y_2)$ (the argument is exactly the same for any TGD, and even simpler for a plain Datalog rule). 
Suppose that atoms 
 $P([c_1]_{\equiv_n},[c_2]_{\equiv_n})$ and $Q([c_3]_{\equiv_n},[c_4]_{\equiv_n})$ are true in  $M_n$.
We need to show that there is an element $[e]_{\equiv_n}\in M_n$ such that 
$R([e]_{\equiv_n},[c_1]_{\equiv_n},[c_2]_{\equiv_n},[c_3]_{\equiv_n},[c_4]_{\equiv_n})$ is also true in $M_n$. 

By definition of $M_n$ there exist elements $a_1, \ldots a_4$ of $Chase$ such that $a_i\equiv_n c_i$ for each $i\in \{1,\ldots 4\}$ and that the atoms $P(a_1,a_2)$ and $Q(a_3,a_4)$ are true in $Chase$. But -- since $Chase$ is a model 
of $\cal T$ --  this means that 
there is an element $e$ of $Chase$ such that the atom $R(e, a_1, a_2, a_3, a_4)$ is also true in $Chase$. 
This however implies that $R([e]_{\equiv_n}, [a_1]_{\equiv_n}, [a_2]_{\equiv_n}, [a_3]_{\equiv_n}, [a_4]_{\equiv_n})$
is true in $M_n$, which is exactly what we needed to prove.\eop

Notice that joinlessness of $\cal T$ was a crucial assumption here. Suppose
 the body of $T$ had the form 
$P(x, v)\;\wedge \;Q(y, v)$ and  atoms $P([c_1]_{\equiv_n},[c]_{\equiv_n})$ and $Q([c_2]_{\equiv_n},[c]_{\equiv_n})$
were true in $M_n$. This still would imply existence 
of atoms $A_1 = P(a_1, a)$ and $A_2 = Q(a_2,a')$, both true in $Chase$, and such that $a_1\equiv_n c_1$, $a_2\equiv_n c_2$
and $a\equiv_n a' \equiv_n c$. But $a$ and $a'$ would not need to be equal, and so rule $T$ could not be applied
to $A_1$ and $A_2$ in $Chase$. This remark explains why -- in order to prove Finite Controllability 
of Sticky Datalog$^\exists$ -- we first reduced the problem to Finite Controllability for Joinless Logic.

Let us also remark that it easy to see that if $\cal T$ is a theory in Guarded Datalog$^\exists$ then 
Lemma \ref{samodelami} remains true. This is why our technique can be directly applied to 
show the FC property  for Guarded  Datalog$^\exists$. Actually, proof in this case is much easier than  in the 
in the case of the Joinless Logic, as the technical details of the proof of Lemma \ref{nfl} significantly simplify.


\definitionnn{\label{eva1} For a conjunctive query $\phi$ let $Occ(\phi)$ be the set of all variable occurrences in $\phi$.
More precisely,
 $Occ(\phi)=\bigcup_{R\in \phi} (\{1,2\ldots arity(R)\}\times  \{R\})$.
	
An $n$-evaluation of $\phi$ is a function $f: Occ(\phi) \rightarrow Chase$ 
assigning, to each atom $R$ from $\phi$ and each position $i$ in $R$, an element $f(i,R)\in Chase$, in such a way that:

 \begin{itemize}
\item[(*)] for each pair of atoms  $R, R'$ in $\phi$ if $R(i)=R'(i')$  then $f(i,R)\equiv_n f(i',R')$.
\item[(**)] for each atom $R$ in $\phi $ it holds that $Chase \models f(R)$. 
\end{itemize}

Where by $f(R)$ we mean the atomic formula resulting from replacing, in $R$, each $R(i)$ 
(which is a variable) by $f(i,R)$ (which is an element of $Chase$).  
}

It is easy to notice that:

\lemmaaa{\label{eva11}
$M_n\models \phi$ if and only if there exists an $n$-evaluation of $\phi$.\eop\vspace{-1mm}
}

See how simple it is: in order to analyze the behavior of queries in the structures $M_n$ we do not need to 
imagine these complicated finite structures at all! The only structure we need to think about is $Chase$,
together with the equivalence relation $\equiv_n$. Imagine a CQ $\phi$ written in the following way.
First there is a conjunction of atoms, and each variable occurs in this conjunction at most once.
Then there is a conjunction of equalities between variables. Of course every CQ can be written like this.
Now, let $\phi'$ be $\phi$ with each equality symbol replaced by $\equiv_n$. What Lemma \ref{eva11} really 
says is that there is no need to ever imagine $M_n$, because 
$M_n\models \phi$ if and only if $Chase \models \phi'$.

To see Lemma \ref{eva11} in action let us now prove the following  two lemmas. We will need them at some point in the future:

\lemmaaa{\label{cicha}
Consider an M-true conjunctive query 
$\phi = PP \wedge R \wedge  \psi $, where $PP$ is a parenthood atom of some
variable $x$ (which means that $PP(1)=x$), and where $R= Q_{F,\delta}(\bar w)$ and $R(i)=x$. 
Let $j<_F i$ be a position in $R$.
Then the position  $\delta(i,j)$  exists in the atom $PP$.}

Notice that if we assumed that $\phi$ is true in $Chase$ then the claim of the Lemma would follow from Lemma \ref{staloscnazw} 
(well, actually it would be Lemma \ref{staloscnazw} then, modulo obvious rewritings): we are already used to the fact that 
if an atom $R$ is true in $Chase$, and there is an argument $a$ in $R$ which calls another argument $b$ ''granny'', then $b$ must occur on the granny position in the parenthood 
atom $PP$ of $a$. 

It is not however immediately clear why the weaker assumption, that $\phi$ is just M-true would be sufficient.\vspace{1mm}

\noindent
{\em Proof of Lemma \ref{cicha}:}
We know that $\phi$ is M-true, so also $M_0\models \phi$. Lemma \ref{eva11} tells us  that there exists a 0-valuation $f$ of $\phi$, which means that
the atoms $f(PP)$, $f(R)$ and $f(P)$, for each atom $P$ in $\psi$, are all true in $Chase$
 and $f$ is such that each two different occurrences of the same variable in $\phi$ are mapped on $0$-equivalent elements of $Chase$. 
So consider the atoms    $f(PP)$ and $f(R)$ in $Chase$. Define  $(f(PP))(1)=a$ and   $(f(R))(i)=a'$.  Since $PP(1)=x$ and $R(i)=x$ we have that  $a \equiv_0 a'$.
Consider the parenthood atom $A$ of $a'$ in $Chase$. By Lemma \ref{staloscnazw} we have that  position  $\delta(i,j)$  exists in $A$. And since $a \equiv_0 a'$
we have that $A$ and $PP$ are atoms of the same predicate, so  position  $\delta(i,j)$ must also exist in $PP$.\eop\vspace{1mm}

\lemmaaa{\label{semantyczny}
Let $\psi$ be an M-true query and let $P_{F,\delta}$ and $R_{G,\gamma}$ be atoms in $\psi$. 
Suppose $x=P(1)=R(j)$, for some variable $x$ and some position $j$ in $R$. Suppose also that
positions $j'$ and $j''$ in $R$ are such that $j'<_G j'' <_G j $. Let $i'$ and $i''$ be such positions
in $P$ that $i'=\gamma(j,j')$ and $i''=\gamma(j,j'')$. Then 
$i'<_F i''  $ and $\delta(i'',i')=\gamma(j'',j')$
}

Notice that positions $i'$ and $i''$ exist in $P$ due to Lemma \ref{cicha}.\vspace{1mm}

\noindent
{\em Proof:} The query $\psi$ is M-true. So consider a 0-evaluation $f$ of $\psi$. Let $a_P=f(1,P)$ and 
$a_R=f(j,R)$. Of course $a_P \equiv_0 a_R$.
Let also $C_P=f(P)$, $C_R=f(R)$ and let $C$ be the parenthood atom of $a_R$
in $Chase$. Of course $C_P$ and $C$ are atoms of the same predicate (because $a_P \equiv_0 a_R$). 

Now use Lemma \ref{staloscnazw} for $a=a_R$, to show that $i'<_F i''  $ and $\delta(i'',i')=\gamma(j'',j')$ hold in 
$C$. This of course implies that they also hold in $C_P$.
\eop 

\section{The second little trick}\label{secondlt}

As we said in Section \ref{in}, for each M-true CQ $\phi$ we will construct its ''normal form'' $\beta$.
The following  lemma describes a single step of the normalization process. Its proof relies on what we find to be 
the  nicest technical idea of this paper\footnote{And explains why the structures $M_n$ are defined as they are.}, so please try to have fun:

\lemmaaa{[Second Little Trick]\label{slt}
Consider an M-true conjunctive query $\phi = P \wedge R \wedge  \psi $, where $P$ is a parenthood atom of some
variable $x$ (which means that $P(1)=x$), and where $R= Q_{F,\delta}(\bar w)$ and $R(i)=x$. 

Let  $\sigma$ be a unification, which for every position $j<_F i$ in $R$
identifies the variable $R(j)$ with the variable $P(\delta(i,j))$ (which exists, due to Lemma \ref{cicha}).
Then $\sigma (\phi) $ is also M-true.
}

Clearly, $\sigma (\phi) $ is more constrained than $\phi$, so whatever structure $\cal M$ we consider
it holds that ${\cal M}\models (\sigma (\phi)\Rightarrow \phi) $ (this observation has something to do with
condition (**) of the Normal Form Lemma).

Notice however that, despite the fact that  $\sigma (\phi) $ appears to be more constrained, we also have:
$Chase \models (\phi \Rightarrow \sigma(\phi))$. This follows from Lemma \ref{staloscnazw}, 
which says that each element -- call it $b$ --  of $Chase$ has a unique tuple of parents, and whenever 
$b=R(i)$ for some atom
$R$  the element $R(j)$ (with $j<_F i$, where $F$, $\delta$ are the family pattern of $R$)
must be the same as the element in position $\delta(i,j)$ in the parenthood atom of $b$. 

This implies that every satisfying  valuation of $\phi$ in $Chase$ must substitute the same element for the variables 
$R(j)$ and $P(\delta(i,j))$
anyway, and so the unification from the Lemma does not really lead to more constraints. 

But the situation in the structures $M_n$ is different. Lemma \ref{staloscnazw} is not valid there, 
as elements of $M_n$ can have more than one tuple of parents. This is because when we identify two $n$-equivalent elements of $Chase$ each of them  
comes with its own parents, and we cannot be sure that the respective 
parents will also be $n$-equivalent, and thus identified. What we know however is that the respective parents will be
at least $(n-1)$-equivalent. And this turns out to be sufficient for:

\noindent {\em Proof of Lemma \ref{slt}:}~
We want to show that for each natural $n$ the query $\sigma(\phi)$ is true in $M_n$.
Fix $n\in \mathbb{N}$. 
We know that $\phi$ is M-true, so $M_{n+1}\models \phi$.

Suppose $f$ is an $(n+1)$-evaluation of $\phi$. The lemma will be proved if we can show that the same function 
$f$ is also an $n$-evaluation of $\sigma(\phi)$. 
Of course condition (**) of Definition \ref{eva1} is still satisfied, as it neither depends on $n$ nor on the equalities between the variables. Also condition (*) is satisfied for the pairs of variables that were already equal in $\phi$.
What remains to be proved is that condition (*) holds true also for  pairs of variables unified by $\sigma$. In other words,
we need to show that 
$f(P(\delta(i,j)),P) \equiv_n f(R(j),R)$
for each position  $j<_F i$ in $R$.

But we know that $ f(P(1),P) \equiv_{n+1} f(R(i),R)$. 
This is because  the variables $P(1)$ and $R(i)$ are equal (to $x$), so $f$, being an $(n+1)$-evaluation, must map them to
elements of $Chase$ which are $(n+1)$-equivalent.
 Since $f$ satisfies condition (**)
of Definition \ref{eva1}, we know (by Lemma \ref{staloscnazw}) that  $f(P(\delta(i,j)),P)$ and $f(R(j),R)$ 
are respective parents of
$ f(P(1),P)$ and $f(R(i),R)$. Now, to end the proof, use the fact that respective parents of $(n+1)$-equivalent
elements of $Chase$ are $n$-equivalent. \eop\vspace{-1mm}


\section{The ordering  ${\rightarrow}_{\phi}$ and cycled  queries}\label{cyclic}

We are already used to the fact that each atom comes with an ordering (''family ordering'') of its arguments.
Now we will extend the (family) ordering on positions of individual atoms to the ordering on variables of conjunctive query the atoms form.
Then we will study  the new ordering very
carefully.

Let us recall that a CQ is clean if it only contains atoms of parenthood predicates. It is also good to remember that 
if $Q_{F,\delta}$ is a parenthood predicate then (the ordering defined by) $F$ is a tree and that position 1 is always the root of this tree.

\definitionnn{
Let  $\phi$ be a clean CQ.

 By  ${\rightarrow}_{\phi}$ we mean the smallest transitive (but not necessarily  reflexive) relation such that
for each $x,y\in Var(\phi)$
if there is an atom $P=Q_{F,\delta}(\bar t)$ in $\phi$ and positions
$i,j$ in $F$,  such that  $P(i)=y$, $P(j)=x$  and $i<_F j$, then $x {\rightarrow}_{\phi} y$.

A CQ $\phi$ is  {\em non-cycled} if ${\rightarrow}_{\phi}$ is a partial order\footnote{
When $x{\rightarrow}_{\phi} y $ then we think that $y$ is smaller than $x$. Mnemonic hint:
the arrowhead of $\rightarrow$ looks like  $>$.
}
 on $Var(\phi)$
(which in particular means that it is antisymmetric). Otherwise it is cycled.

}

Clearly, if $\phi$ is cycled then $Chase\not\models \phi$. But it is also not hard  to see that:

\lemmaaa{\label{cykliczne}
 If $\phi$ is a cycled query consisting of $k$ atoms, then $M_{k+1}\not\models \phi$. So a cycled query
 is never $M$-true. 
}

\noindent
{\em Proof:} Let sequence of atoms $R_1, \ldots R_{j-1}$ for $j\leq k$ be a witness of the fact that $\Phi$ ic cycled. 
It means that there exist a sequence of variables $x_1,\ldots x_j$ such that $x_{i}$ is a parent of $x_{i+1}$ in atom $R_i$ and $x_1=x_j$.

Suppose that $\phi$ is true in $M_k$ (and therefore in $M_j$). Let $f$ be a $j$-evaluation of $\phi$. 
From $f$ we can extract two sequences $a_1,\ldots, a_{j-1}$ and $b_2,\ldots, b_{j} $ such that
\begin{itemize}
 \item $a_i$ is a parent of $b_{i+1}$ in $Chase$ 
 \item $\forall _{1<i<j}~~a_i\equiv_j b_i$
 \item $a_1\equiv_j b_j$
\end{itemize}
By abusing the notation a bit we could just say that $a_i=f(x_i,R_i)$ and $b_{i+1}=f(x_{i+1},R_i)$.

\begin{observ}There exists  a sequence $c_1, \ldots, c_j$ of elements of $Chase$ such that
\begin{itemize}
 \item $c_i$ is a parent of $c_{i+1}$ in $Chase$
 \item $\forall _{i<j}~~ c_i\simeq^{j}_{i}a_i$
 \item $c_j=b_j$
\end{itemize}
\end{observ}

Notice that once the Observation is proved, the proof of Lemma \ref{cykliczne} will be finished: 
this is because it follows from the Observation that  $c_1 \simeq^{j}_{1} a_1\equiv_j b_j =c_j$, which means that $c_1$ has the same $j$-color as $c_j$.
But this leads to a contradiction since $c_1$ is in a $j$-history of $c_j$, and this is exactly what is prohibited by Definition \ref{coloring}.

\noindent
{\em Proof of Observation:} This sequence will be constructed by induction. Let $c_j=b_j$ and $c_{j-1}=a_{j-1}$.

Suppose that $c_{i+1}$ has been defined. Since $c_{i+1}\simeq^{j}_{i+1}a_{i+1} \equiv_j b_{i+1}$ we have $c_{i+1}\simeq^{j}_{i+1} b_{i+1}$. 
Because $a_{i}$ is a parent of $b_{i+1}$, there must exist a respective parent $c_i$ of $c_{i+1}$   such that $c_{i}\simeq^{j}_{i} a_{i}$. \eop\vspace{1mm}

This was fortunately the last time we needed to think about colors.

It follows from Lemma \ref{cykliczne} that in the proof of the Normal Form Lemma (Lemma \ref{nfl}) we only need to consider non-cycled queries.


\section{Non-cycled queries and the normal form}\label{nfd}

 Now please be ready for the most technical part of the paper.
 Let $\phi$ be an non-cycled and M-true CQ and let 
 ${\rightarrow}_{\phi}$ be the partial order on 
$Var(\phi)$, as defined in the previous section.

\definitionnn{
 Call a variable $x\in Var(\phi)$ {\em important} if $x=P(1)$ for some  atom\footnote{Do not forget that
only parenthood atoms appear in queries} $P$ in $\phi$.
Otherwise $x$ is called ordinary.}

So the important variables are the ones we know a lot about -- we know all their
parents by name.

%

Let us remind the reader that the notation $PY$ was introduced in  Definition \ref{py}.

\definitionnn{\label{skomplikowana}
\begin{itemize}
\item
For an atom $P=Q_{F,\delta}(\bar t)$ of $\phi$ let ${\cal I}(P)$ denote the 
 set of such 
non-root positions $i$ in $P$ that the variable $P(i)$ is important and that for each $j\neq 1$ if $i<_F j$ then
$P(j)$ is ordinary.

\item
For an atom $P$ of $\phi$
define $top.pos(P)= PY({\cal I}(P))$.
Let $top.pos(\phi)\subseteq Occ(\phi)$ be the set of such variable positions $(i,P)$  that 
 $i\in top.pos(P)$. 

 \item
For an atom $P$ of $\phi$ let $top.var(P)=\{P(j):j\in top.pos(P)\}$. A variable $y\in Var(\phi)$ is  a top variable
if $y\in top.var(P)$ for some atom $P$ of $\phi$.

\end{itemize}
}

In other words 
$top.pos(\phi)$ is the set of  positions in the atoms of $\phi$, which are, in a certain sense 
''close to the roots''  of the respective atoms --
there are no important variables between this position and the root of the atom. The set $top.var(P)$ is a set 
of variables -- these variables that occur in one of the ''top positions'' of $P$.
  
Now we can define the normal form of a conjunctive query:

\definitionnn{\label{dnf}
A CQ $ \phi$ is in the normal form if:

\begin{enumerate}
\item[~]\hspace{-6mm}{\em Ideological condition:~}  
If $P$ is an atom in $\phi$ which is a  parenthood atom of an important variable $x$, if $R=Q_{F,\delta}(\bar t)$ is another 
atom in $\phi$, such that  $R(i)=x$, and if $j$ is a position in $R$ such that $j<_F i$, then $R(j)=P(\delta(i,j))$.

\item[~]\hspace{-6mm}{\em Technical condition:~} Each variable from $Var(\phi)$ occurs in at most one
position in $top.pos(\phi)$. 

\end{enumerate}
}

Notice that it follows from the Ideological Condition, that an important variable $x$ of a query $\phi$ in the normal form 
can be in the root position in only one atom of $\phi $ (a query is a set of atoms, so 
equal atoms count as one). In order to see that suppose that there are two such atoms, $P$ and $R$. Since $\phi$ is assumed to be $M$-true, $P$ and $R$ 
must be atoms of the same predicate. Now apply the Ideological Condition to $P$ and $R$ and see that it follows that variables in the same positions
in $P$ and $R$ must be equal, so $P$ and $R$ are in fact one atom. 
Call this unique atom having $x$ in the root position $PP_x$.

 Since the root positions are the only positions of 
important variables which are 
in $top.pos(\phi)$ this means that the 
Technical Condition for the important variables is implied by the Ideological Condition.

Notice also  that the Ideological Condition 
 is the condition from Lemma \ref{slt}.
So one can imagine now, how we are going to prove Lemma \ref{nfl} -- we will start from the query $\phi$
(or from something  similar -- actually it is not going to be exactly $\phi$) and  perform the 
unifications from Lemma \ref{slt} on it, as long as possible.
 The main difficulty 
in the proof of Lemma \ref{nfl} is to make sure that the final result of 
such a  unification procedure
 indeed satisfies the Technical Condition for the ordinary variables, which 
 will be very much needed  (in Section \ref{ch}) for the proof of the Lifting Lemma.

For the (mostly boring and syntactical) details of the proof of Lemma \ref{nfl} see the next two sections.  

As it turns out, the assumption that  an M-true query $\phi$ is in the normal form, or even that
 it satisfies the Ideological Condition alone, implies a lot  about the ordering $\mlodszy{\phi}$:

\definitionnn{Let $y,y'\in Var(\phi)$. We will call $y'$ a successor of $y$ if 
$y'\mlodszy{\phi}y$ and there is no such $z\in Var(\phi)$ that $y'\mlodszy{\phi}z$ and $z\mlodszy{\phi}y$.
}

\lemmaaa{\label{zideologii}
Let $\phi$ be an non-cycled M-true query satisfying the Ideological Condition. 
Then:
\begin{enumerate}
\item[A.]
 Every variable in $Var(\phi)$ is a top variable.
\item[B.]
 If an ordinary variable $y'$ is a successor of an ordinary variable $y$ then there is an atom $PP_x$
such that $y,y'\in top.var(PP_x)$. If an important variable $x$ is a successor of an ordinary variable $y$ then 
$y\in top.var(PP_x)$.
\end{enumerate}
}

\noindent
{\em Proof of A:}
Suppose there is a variable $y\in Var(\phi)$ which is not a top variable. Let $z$ be a minimal,
with respect to the ordering $\mlodszy{\phi}$, important variable such that $y\in Var(PP_z)$.
Let $<_F, \delta$ be the family pattern of $PP_z$.

We know that $y\not\in top.var(PP_z)$, so there must be an important variable $x\in  Var(PP_z)$
such that $x\neq z$ and $i <_F j $, where 
 $PP_z(i)= y$ and $PP_z(j)= x$. But 
this means, since $\phi$ satisfies the Ideological Condition, that $y$ occurs in 
the atom $PP_x$ (in position $\delta(j,i)$), which contradicts the minimality of $z$. 

Notice that we silently used Lemma \ref{cicha} here, and this is where the assumption that $\phi$ is 
M-true was needed.\vspace{1mm}

\noindent
{\em Proof of B:}
If $y'$ (ordinary or important) is a successor of $y$ then, by the definition of $\mlodszy{\phi}$,
there must be an atom $PP_x$, with the family ordering $<_F$,
and positions $i,i'$ in $PP_x$,  such that 
 $i <_F i'$, $PP_x(i)=y$, $PP_x(i')=y'$.
Notice also that, if $i$ and $i'$ are as above, there is no position $j$ satisfying 
$i <_F j  <_F i'$ -- this is because the variable $PP_x(j)$  would be between $y$ and $y'$ in the ordering 
$\mlodszy{\phi}$. Let $x$ be a minimal,
with respect to the ordering $\mlodszy{\phi}$  variable such that $PP_x$ satisfies the above requirements.
Now, use the argument from the proof of claim A. to show that  $i$ is a top position in $PP_x$. \eop

\lemmaaa{\label{ztechniki}
Let $\phi$ be an non-cycled M-true query in the normal form. Then:
\begin{enumerate}
\item[A.]
Each ordinary variable has exactly one successor.

\item[B.]
Suppose $y\in top.var(PP_x)$, the variable $z$ is important and $z\mlodszy{\phi} y$. Then 
$z\mlodszy{\phi} x$.
\end{enumerate}
}

\noindent
{\em Proof of Lemma \ref{ztechniki}.} Claim A. follows directly from Lemma \ref{zideologii}B and from 
the Technical Condition. Claim B. follows directly from A. \eop\vspace{1mm}

Now all the notions appearing in the Normal Form Lemma and in the Lifting Lemma are defined and what remains to be done is proving the two Lemmas.
The next two sections are devoted to the proof of Lemma \ref{nfl}. But once you know the main idea,  which is performing the unifications from the Second Little Trick
as long as needed/possible, the proof is hardly exciting. Then, in the last section of the paper, the Lifting Lemma (Lemma \ref{chl}) is proved, and this is where the rabbit is pulled out 
of the hat. So maybe it is not a bad idea to skip Sections \ref{nudna01} and \ref{nudastraszna} and jump directly to Section \ref{ch}. 


\section{Proof of Lemma \ref{nfl}. Part one: the normal form of $\phi$.}\label{nudna01}

In this Section we consider some fixed M-true CQ $\phi$ and  construct a CQ $\beta$ being the normal form
of $\phi$, as specified by Lemma \ref{nfl} and Definition \ref{dnf}.

The definition  of $\beta$  itself (Definition \ref{beta}) is quite natural and not very complicated.
The really technical part  begins right after Definition \ref{beta}, where we prove that 
the defined query is indeed the normal form of $\phi$. There are no deep ideas there, we just need to
carefully analyze the consequences of the unifications resulting from applications of the Second Little Trick,
and such analysis is, by its nature, a very syntactic thing.\vspace{1mm}

\noindent
{\sc notational conventions.} 
The typical situation in this part of the paper will be that we will consider some {\bf fixed}
CQ $\theta$, and restrict attention only to
queries being equality variants of $\theta$. By this we mean queries that can be obtained from $\theta$ by renaming 
some of the occurrences of variables. 

We need a convenient language for this scenario, so let us start from defining such a language.

Equality  variants of $\theta$ only differ by the names of the variables, and they all have the same
set of positions. 
We will imagine that $\theta$ is a conjunction of some 
atoms $P^l_{F_l,\delta_l}$, with  $l\in V$ for some set $V$, and we will denote
by $\cal P$ the set of all positions in $\theta$ 
(which is a disjoint union of the sets of positions in the atoms). By saying ''let $i\in \calp$'' 
 we can now address a position directly,
without specifying in which of the atoms of $\theta$  it is located. The cost to pay is that no longer we can
use 1 for the name of the position in the root of the atom, so by $root(i)$ we  will mean that
$i\in \calp$ is a position in the root of some $P^l$. By $\calp_\xi(i)$ (or just  $\calp(i)$ when the context is clear) 
we will mean the variable in position $i\in \calp$ in the equality variant $\xi$ of   $\theta$.

Let $\prec$ be the disjoint union of all relations $<_{F_l}$, 
so that by $i\prec j$, for $i,j\in \cal P$,  we mean that positions $i$ and $j$ are in the same 
atom $P^l$, for some $l$, and $i<_{F_l} j$. Similarly, let $\delta$ be the disjoint union of all the functions $\delta_l$.
 
It will be also convenient to have a notation 
$\Delta(i,i,j',j')$ for the formula $root(i) \;\wedge \; \delta(i,i') = \delta(j,j')$.

In other words (for those who do not like our new language) $\Delta(i,i,j',j')$ means that 
there are $l$ and $l'$ such that $i$ is the position in the root of $P^l$, $i'$ is a position in $P^l$,
$j$ and $j'$ are positions in $P^{l'}$,  and $\delta_l(i,i') = \delta_{l'}(j,j')$.

Since the objects defined in this subsection ($\cal P$, $\Delta$, $\prec$, $\delta$)  depend on our current
choice of $\theta$, they only have meaning  in the contexts where $\theta$ is defined.

See how conveniently the Ideological Condition from Definition \ref{dnf} can  now be expressed:\\

\noindent
($\heartsuit$)~ for each $i,i',j,j'\in \calp$, if $\Delta(i,i',j,j')$  and $\calp(i)=\calp(j)$ then $\calp(i')=\calp(j')$.\\


\noindent
{\sc The unification procedure.} For a query $\psi$ let
 $u(\psi)$ be a result of:

\noindent
{\bf The unification procedure:}\\ 
fix $\theta$ as $\psi$;\\
 \hspace*{5mm}$/^*$  So that the above notations apply $*/$ \\
$\xi:=\psi$\\
{\bf while} there exist:
$i,j,i',j'\in \calp$ such that $\Delta(i,i',j,j')$, ~~ $\calp_\xi(i)=\calp_\xi(j)$ and $\calp_\xi(i')\neq \calp_\xi(j')$\\
{\bf do}\\
$\{$\\ replace all occurrences of  $\calp_\xi(j')$ in $\xi$ by $\calp_\xi(i')$\\
(in other words $\xi:=\xi [\calp_\xi(j')/ \calp_\xi(i')]$);\\
$\}$\\
forget that $\theta$ was $\psi$;\\
\hspace*{5mm}$/^*$  So that we can use $\theta$  somewhere else  $*/$ \\
remove the repeating atoms from $\xi$;\\
{\bf return} $\xi$ as $u(\psi)$;\\
{\bf end of the unification procedure.}\\

What this procedure does is exactly checking if the Ideological Condition is satisfied in $\xi$, and
if it isn't, unifying the variables that violate the Ideological Condition, using the Second Little Trick.
Clearly, the procedure always terminates and $u(\psi)$  always satisfies the Ideological Condition.
We also know, from Lemma \ref{slt}, that if $\psi$ is M-true then $u(\psi)$ also is. It is also obvious that
$Chase\models (u(\psi)\Rightarrow \psi)$. 

We are however not claiming that $u(\psi)$ is always the normal form of $\psi$. This is because there is no reason
for the Technical Condition  to be satisfied in $u(\psi)$. One could for example easily take $\psi$ to be a
query which already satisfies the Ideological Condition (so that $u(\psi)=\psi$) but not the  Technical Condition. 

The unification procedure is nondeterministic -- at each step it nondeterministically selects, for the unification,
a pair of variables. But:

\lemmaaa{\label{fixpoint}
The result of the unification procedure -- the $u(\psi)$ -- is  unique for $\psi$,
in the sense that it does not depend on the nondeterministic choices made by the procedure.}
 
\noindent
{\em Proof:} Since the set of positions $\calp$ is fixed, a query $\xi$ can be identified with its equality relation 
$ =_\xi $ on the set of positions (this relation says that the variables in two positions are equal in $\xi$).
 What the unification procedure does is computing the fixpoint of some Datalog program. The relations 
 $\Delta$ and $=_\psi$ are the input predicates of this program
 while $=_{u(\psi)}$ is its output predicate. The rules of the program are the condition $(\heartsuit)$ above, and 
the reflexivity, symmetricity and transitivity axioms for $=_{u(\psi)}$. 
And of course the fixpoint of a Datalog program does not depend on the order of execution.  \eop\\


\noindent
{\sc elevating the importance of the variables.}
As we said, $u(\psi)$ is not always in the normal form, as it may not satisfy the Technical Condition. The Technical 
Condition concerns the ordinary variables, and the reason why $\psi$ may not satisfy it is that there may be some
unwelcome equalities between ordinary variables in $\psi$. Our way towards the solution of the problem is 
 to elevate  the (potentially) misbehaving ordinary variables to the position of importance,
so that they are allowed more. 

\definitionnn{
For a query $\psi $ by a closure of $\psi$ we will mean any query of the form
$\psi \wedge \bigwedge_{x\in Var_{ord}(\psi)} R(x,\bar x)  $
where $Var_{ord}(\psi)$ is the set of all the ordinary variables of $\psi$, $R$ is any parenthood predicate and 
$\bar x$ is a tuple of fresh variables.  
} 

It is now straightforward to see that:

\lemmaaa{
 if $\psi'$ is any closure of $\psi$ then:
\begin{itemize}
\item
 if $x\in Var(\psi)$ then $x$ is important in $\psi'$;
\item
 each ordinary variable in $\psi'$ occurs in $\psi'$ only once;
\item
$Chase\models (\psi' \Rightarrow \psi)$;
\item
$\psi'$ satisfies the Technical Conditions (although not necessarily the Ideological Condition).
\end{itemize}
}

It is also not hard to show that:

\lemmaaa{
If $\psi$ is M-true then there exists an M-true  $\psi'$ being a closure of $\psi$.
}

\noindent
{\em Proof:} For each $n\in \mathbb{N}$  if $M_n\models \psi$ then also  $M_n\models \psi$
 for some closure $\psi'$ of $\psi$. This is because each element of $M_n$ is a child in some 
parenthood atom valid in $M_n$. 

Since there are only finitely many possible closures of $\psi$, if $\psi$ is M-true, then there is a closure 
$\psi'$ which is true in $M_n$ for infinitely many numbers $n$. Now use the argument from the First Little Trick. \eop

From now on, for an $M$-true conjunctive query  $\psi$ by  $c(\psi)$ we will denote an  M-true closure of $\psi$.\\


\noindent
{\sc the query $\beta$ -- the normal form of $\phi$.}
We are finally ready to name the query $\beta$ which is the normal form of $\phi$:

\definitionnn{\label{beta}
 $\beta = u(c(\phi))$.
}

\lemmaaa{\label{nuda}

\begin{enumerate}
\item 
 $ \beta$ is M-true;
\item 
 $ Chase\models (\beta \Rightarrow \phi) $;
\item
 $\beta $ satisfies the Ideological Condition;
\item 
 $\beta$ satisfies  the  Technical Condition.
\end{enumerate} 
}

Claims 1)--3) follow immediately from the construction. But Claim 4) is not obvious at all, it needs a proof, and this
proof, while not really complicated,  is unfortunately not going to be short. 
Notice however  that once Lemma \ref{nuda} is proved then of course also the proof of Lemma \ref{nfl} will be finished.


\section{Proof of Lemma \ref{nfl}. Part two:
proof of Lemma \ref{nuda}.4.}\label{nudastraszna}

 What remains to be done to show that  $\beta$ is indeed the normal form of $\phi$ is proving that 
it satisfies  the  Technical Condition.
The main proof technique is a patient syntactical analysis of the unifications that led to $\beta$.

Let now $\theta$ -- the query with respect to which the notations are defined in the beginning of the previous Section -- be equal to $\beta$. And this is not going to change any more. 

Before we show Lemma \ref{nuda} let us try to imagine how $\beta$  looks like. 
There are two kinds of atoms in $\beta$. One are those that originated in $\phi$. Now they contain only important variables.
Second kind are the atoms that were originally added to $\phi$ when $c(\phi)$ was created. They may contain ordinary variables,
but also, after all the unifications on $c(\phi)$ they contain some important variables in non-root positions.\\\vspace{1mm}

\noindent
 {\em Proof of Lemma \ref{nuda}.4.} 
Please allocate memory for two more equality variants of $\beta$. They will be called $\beta_0$ and
$\beta_{wu}$ (as ''weakly unified''), which will at the end turn out to  actually be equal to $\beta$.

We need to do something strange now. 
Due to a  reason  that  will be explained later, we need to destroy the structure of $\beta$, to some
extend, and then to rebuild it again:

\definitionnn{
Let  $\beta_0$ be the result of substituting a fresh variable for each occurrence of an ordinary variable in  
$\beta$.}

Of course $\beta_0 $ is not simply $c(\phi)$.
The unifying procedure run on $c(\phi)$ {\bf a)} unified some of the fresh variables in the new atoms of $c(\phi)$  with the 
 variables  from $Var(\phi))$, and
{\bf b)} unified some of these  fresh variables with other fresh variables. The query $\beta_0$ is the result
of undoing the unifications from {\bf b)}, but not from {\bf a) }. 

\lemmaaa{
%

$u(\beta_0)= \beta $;

}

This is because  $\beta=u(c(\phi)) $ is more unified than 
$\beta_0$ and $\beta_0$ is more unified than $c(\phi)$. Use the datalog fixpoint argument from the proof of Lemma \ref{fixpoint}.\eop

Clearly, $\beta_0$ satisfies the Technical Condition. 

Now we are going to run a version of the unification procedure on $\beta_0$,
which will lead us to a new query $\beta_{wu}$. The query $\beta_{wu}$ is in fact $\beta$,
 but this is a secret yet.
In this new unification procedure the
 pairs of variables to be unified, will be carefully hand-picked in the {\em correct order} 
and nothing will be left to
nondeterminism. Thanks to that
we will be able to make sure that the Technical Condition keeps being satisfied. 
One of course could  ask here why did we bother to define $\beta $ first, 
if then we run another unification procedure on $\beta_0$  anyway?
And the answer is, that we only can know the correct order once we know $\beta$! So we need to know $\beta$,
constructed in any way, to be able to construct $\beta$ again in the careful way. 

Notice that whatever our order of the execution of the unification procedure is going to be, 
 we will never unify any important variable with any other variable (important or ordinary). 
If $x$ is an important variable in $\beta$ then it is also important in $\beta_0$ and for 
each $i\in \calp$ we have that $\calp_\beta(i)=x$ if and only if $\calp_{\beta_0}(i)=x$. This observation leads 
to a series of definitions:

\definitionnn{\label{rodzynki}
Call a position $j\in \calp$  {\em ordinary}, if the variable  $\calp_{\beta_0}(j)$ is ordinary (or -- equivalently --
if the variable  $\calp_\beta(j)$ is ordinary).
Otherwise $j$ is important. Let $\calp_{ord}$ and  $\calp_{imp}$ denote, respectively,
 the sets of ordinary and important positions.
}

\definitionnn{
For an ordinary position $j\in \calp$  denote by $nearest.pos(j)$ the smallest, with respect to 
the ordering $\prec$, important position $i$ in $\calp$ such that $ j \prec i $.
By $nearest.var(j)$ denote the variable $\calp(nearest.pos(j))$. 
}

In other words $nearest.pos(j)$ is the first important position on the path 
from $j$ to the root of the atom where $j$ is located, and $nearest.var(j)$ is the name of the important variable that
lives there. 

\definitionnn{~\\
For an important variable $x$ let
$layer(x)=\{j\in\calp_{ord}: nearest.var(j)=x\}.$
}

Of course:

\lemmaaa{
The sets $layer(x)$, for $x\in Var_{imp}(\beta)$, form a partition of $\calp_{ord}$ (by which we mean that
they are pairwise disjoint and that their union equals $\calp_{ord}$).
}

Let us also remind that an ordinary position $j$ is a top position if $root(nearest.pos(j))$ 
(this is Definition \ref{skomplikowana} in our new language).


Now we are ready for:\\

\noindent
{\bf The weak unification procedure:}\\
$\xi := \beta_0 $;\\
{\em to-be-considered} $:= Var_{imp}(\beta_0)$\\

\noindent
{\bf while} {\em to-be-considered}$\neq \emptyset$\\
{\bf do:}\\
\{ \hfill $/^*$ $\diamondsuit$ $*/$ \\ 
Let $x$ be a minimal, with respect to the ordering $\rightarrow_\beta$, variable in the set {\em to-be-considered};\\
\hspace*{4mm}$/^*$ See! Here is where we need to know $\beta$. \hfill  $*/$ \\
Let $i\in \calp $ be such that $\calp(i)=x$ and $root(i)$;\\
\hspace*{4mm}$/^*$ We took the position in the root of the atom $PP^x$. \hfill $*/$ \\
 For each non-top position $j'$ such that $j' \in layer(x)$,\\ 
\hspace*{2mm}and for each $i'$ such that $\Delta(i,i',nearest.pos(j'),j')$\\
\hspace*{4mm}substitute the variable $\calp(j')$ in $\xi$ by the variable $\calp(i')$;\\
\hspace*{4mm}$/^*$ Call the above the ''unification step'' \hfill $*/$ \\
Remove the variable $x$ from {\em to-be-considered};\\
\}\\
Return $\xi$ as $\beta_{wu}$.\\
{\bf end of the procedure.}\\

Let us try to explain the substitution step of the procedure. 

Once $x$ is fixed (which is one of the   $\rightarrow_\beta$ minimal variables not yet considered) we look for all possible
positions $j'\in \calp_{ord}$, such that the if we started, in $j'$, a path (in the ordering $\prec$) towards the root
of the atom where $j'$ is located, the first important position on this path would be some non-root
position $j=nearest.pos(j')$, and the variable there would be $x$. 

Then we ask $j$: ''how do you call $j'$ ?''. And we get some answer ''$\delta(j,j')$''. So we ask $i$: ''whom do you call 
$\delta(j,j')$ ?''. And we get some answer ''$i'$''. Then we say: ''So, since the variables in $i$ and $j$ are equal, the Ideological Condition wants the variables in
$i'$ and $j'$ to unify. From now on the one in $j'$ will adopt the name of the one in $i'$''.

Of course unification means more than  just renaming the variable in $j'$. We need to rename all the 
occurrences of $\calp(j')$ in the current $\xi$. But the trick is that:

\lemmaaa{
Each time the control passes the point marked with $\diamondsuit$, 
if  $x\in$ {\em to-be-considered} and $j\in layer(x)$ then $\calp(j)$ is a fresh variable (which means that it only 
occurs once in $\xi$).
}   

\noindent
{\em Proof:} There are two ways for a variable to lose its freshness. One is to be copied somewhere, which means
being the $i'$ from the unification step,
 the other is
to be substituted with another variable, which means being the $j'$ from the unification step.

But notice that each non-top position in $\cal P$ is exactly once the $j'$ from the unification step, and right after that 
the variable $nearest.var(j')$ is removed from  the set {\em to-be-considered}. Notice also, that
each position that, at some point of time, had already been the $i'$ of the unification step,
must be a position in some atom $PP^z$, with $z$ not being in the set   {\em to-be-considered} any more 
(because in the unification step we take the names for the variables from the atom having the currently considered variable $x$
in the root).  And if $k\in layer(x)$ and $x\in$ {\em to-be-considered} then $k$ is  a position 
in the atom $PP^z$ for some $z$ such that $z\mlodszy{\beta} x$, which implies that $z\in $ {\em to-be-considered}.\eop

The meaning of the last lemma is that the substitution in the unification step is just a renaming of 
one variable occurrence -- the one in $j'$. It does not propagate, in the sense that it does not force any other renamings. 
This means that there is just one chance for a position, during the execution of the procedure, to have its
variable changed -- when this position is the $j'$ from the unification step. Since only non-top positions 
are ever the $j'$, the next lemma follows:

\lemmaaa{\label{technicznywarunek}
If $j$ is a top position in $\calp$ then $\calp_{\beta_0}(j) = \calp_{\beta_{wu}}(j)$
}

Lemma \ref{technicznywarunek} implies that the query $\beta_{wu}$ satisfies the Technical Condition.
But we still cannot be sure that it also satisfies the Ideological Condition. While 
the while loop from the original unification procedure (from Section \ref{nudna01})
 really checks for the premise of the Ideological Condition and,
if this premise  holds, 
it performs the unifications, and does it as long as needed, the loop in the weak unification procedure 
only performs some hand-picked unifications. We need one more lemma to improve our understanding of 
how the query $\beta_{wu}$ looks like:

\lemmaaa{\label{przedostatni} If, at some point of the execution of the weak unification 
procedure, the variables in positions $i'$ and $j'$ were unified (i.e. the variable from 
$i'$ was copied to $j'$) then they remain equal  in $\beta_{wu}$
}

\noindent
{\em Proof:} As we said before, the variable in each position can only be changed once by the weak unification procedure.
So the variable in $j'$ will not be changed any more. We need to make sure that the variable in $i'$ will not be changed
after it was copied to $j'$. 
Suppose the variable $x$ was being considered when the variables in positions $i'$ and $j'$ were unified.
This means that either $i'$ is a top position in $PP_x$ (which means, as we observed before,
 that the variable there can never be changed) or $i'\in layer(z)$ for some $z$ such that $x\mlodszy{\beta} z$.
But this means that at the moment of the unification $z$ is no longer in the set  {\em to-be-considered}, 
and so the variable in $i'$ was already substituted, and it never will again.
\eop

Now the last lemma we need to show in 
order to finish the proof of Lemma \ref{nuda}:

\lemmaaa{
The query $\beta_{wu}$ satisfies the Ideological Condition. In consequence, $\beta_{wu}=\beta$.
}

\noindent
{\em Proof:} We know from Lemma \ref{przedostatni} that $\beta_{wu}$ is weakly unified, which means that if 
$i,i',j,j'$ are positions in $\cal P$ such that $\Delta(i,i'j,j')$, if $\calpbu(i)=\calpbu(j)$,
and if $j=nearest.pos(j')$ then $\calpbu(i')=\calpbu(j')$.

What we need to show is that the Ideological Condition holds, that is 
 if $i,i',j,j'$ are positions in $\cal P$ such that $\Delta(i,i'j,j')$, if $\calpbu(i)=\calpbu(j)$,
 then $\calpbu(i')=\calpbu(j')$.

Suppose that the above is not true and let $x$ be a minimal, with respect to the ordering $\mlodszy{\beta}$,
important variable such that there exist positions $i,i',j,j'$  in $\cal P$ such that $\Delta(i,i'j,j')$ and  $\calpbu(i)=\calpbu(j)$ but
$\calpbu(i')\neq \calpbu(j')$.

Let $y$ be an important variable such that $j'\in layer(y)$, and let $k_j=nearest.pos(j')$ 
(so that $\calpbu(k_j)=y$). Of course it cannot be that $k_j=j$, as this would contradict the assumption that
$\beta_{wu}$ was weakly unified. So we have $j'\prec k_j \prec j$. 

Let $k_i\prec i$ be such position that $\delta(i,k_i)=\delta(j,k_j)$.
From Lemma \ref{semantyczny} we know that $i'\prec k_i$ and $\delta(k_i,i')=\delta(k_j,j')$. 

Notice that $\delta(i,k_i)=\delta(j,k_j)$ implies that $\calp_\beta(k_i)=  \calp_\beta(k_j)$. This is
because the variables in $i$ and $j$ are equal in $\beta$ and $\beta$ satisfies the Ideological Condition.
But $\calp_\beta(k_i)=  \calp_\beta(k_j)=y$ is an important variable, so we have that 
$\calpbu(k_i)= \calpbu(k_j)=y$.  

Let now $k\in \calp$ be such that $root(k)$ and $\calpbu(k)=y$. Such $k$ must exist because 
each important variable is a root somewhere. Let $k'$ be such that $\delta(k,k')=\delta(k_j,j')$ 
(and thus also $\delta(k,k')=\delta(k_i,i')$).

 Since $x\mlodszy{\beta} y$, by the minimality of $x$ we now get
 that $\calpbu(k')=\calpbu(j')$ and $\calpbu(k')=\calpbu(i')$. Contradiction.\eop

This ends the proof of Lemma \ref{nuda} and of Lemma \ref{nfl}.


\section{Proof of the Lifting Lemma}\label{ch}
In this section we  show what remains to be shown: 
 that  if  $M_0 \models \psi$ and $\psi$ is in the normal form then also $Chase \models \psi$. 

As we remember from Section \ref{mn},  $M_0 \models \psi$ means that there exists a 0-evaluation of $\psi$. Such a 
0-evaluation is a function assigning to each variable occurrence in $\psi$ an element of $Chase$ in such a way that 
the atoms in $\psi$ map into atoms true in $Chase$ and (different occurrences of) equal variables map to
0-equivalent elements of $Chase$.   $Chase \models \psi$ means almost the same, the only difference is that 
equal variables map to equal elements of $Chase$, not just to 0-equivalent.

\definitionnn{\label{eva}
A 0-evaluation $f$ is faithful with respect to a set $S\subseteq Var(\psi)$ if 
for each pair of atoms  $R, P$ in $\psi$ such that $Var(R),Var(P) \subseteq S$
if $R(i)=P(i')$  then $f(i,R) = f(i',P)$
}

If $f$ is faithful with respect to $S$ then
for an atom $R$ in $\psi$, such that $Var(R)\subseteq S$, and for $z=R(i)$,  we write $f(z)$ instead of $f(i,R)$. 
 
Being faithful with respect to $S$ means to look, inside $S$ like a real valuation of a $\psi$ in $Chase$.
Clearly
 $Chase\models \psi$ if and only if there exists  a  0-evaluation faithful with respect to $Var(\psi)$. On the other hand,
since 
 $M_0\models \psi$, we know that  there exists  a  0-evaluation faithful with respect to $\emptyset$. 
 We are going to gradually modify  this
0-evaluation to make it more and more faithful, until we get  one faithful with respect to $Var(\psi)$.

The sets $S$ we will be interested in are ideals in $Var(\psi)$:

\definitionnn{\label{idealy}
	Subset $S\subseteq Var(\psi)$ is an important ideal if:
 \begin{enumerate}
\item If $x\in S$ and $x {\rightarrow}_{\psi} y$ then also $y\in S$.
\item All maximal elements of $S$  are important variables.
\end{enumerate}
}

\noindent
{\bf From now on} let $S$ be an important ideal and let $x\in Var(\psi)$
be a minimal important variable not in $S$. Let $PP_x=Q_{F, \delta}(x,\bar x)$
 be, as usually, the parenthood atom of $x$ in $\psi$.
Let $S'$ be the important ideal generated by $x$ and $S$.

\lemmaaa{\label{chwilkaa}
\begin{enumerate}
\item If $R$ is an atom in $\psi$ such that $Var(R)\subseteq S'$ but $Var(R)\not\subseteq S$ then $R=PP_x$.

\item $S'\setminus S = top.var(PP_x) $
 \end{enumerate}
}

\noindent
{\em Proof:} 1) Each atom in $\psi$ is the PP atom of some important variable. If $R$ is the PP atom of some
$y\in S$ then $Var(R)\subseteq S$. If $R$ is the PP atom of some
$y\not\in S'$ then of course  $Var(R)\not\subseteq S'$. And $x$ is the only important variable in  $S'\setminus S$.

2) This follows easily from Lemmas \ref{zideologii} and \ref{ztechniki}. Let us show, for example, that  
$top.var(PP_x) \subseteq S'\setminus S $. Of course $top.var(PP_x) \subseteq S' $ so what we need to
show is that $top.var(PP_x) \cap S =\emptyset $. Let $y\in top.var(PP_x)$. Suppose $y\in S$. This would mean
that there exists an important $z\in S$ such that $z\mlodszy{\psi} y$. But, by Lemma \ref{ztechniki}, this would 
imply that  $z\mlodszy{\psi} x$, which is a contradiction. The proof of the other inclusion is left as an easy exercise. \eop

We will need the following easy observation about local (restricted to one atom only) modifications of  0-evaluations:

\definitionnn{\label{podobne}
Suppose $f$ is a 0-evaluation, $f': Occ(\psi) \rightarrow Chase $ is any function,
and $P$ is an atom in $\psi$.
We say that
$f'$ is $P$-similar to $f$ if:

\begin{itemize} 
\item $f'(i,R) = f(i,R)$  for each atom $ R\neq P$, and each position $i$ in $R$;
\item $Chase\models f'(P)$
\item $f'(i,P) \equiv_0  f(i,P)$ for each position $i$ in $P$.

\end{itemize}
}

\lemmaaa{\label{chwilkab}
If $f$ is a 0-evaluation and  $f'$ is $P$-similar to $f$
then $f'$ is also a 0-evaluation. \eop
}

Let $S,S'$ and $x$ be as  above. In view of Lemma \ref{chwilkaa} 1) and Lemma \ref{chwilkab},
due to an induction argument, in order to prove  Lemma \ref{chl}, it now only remains to show:

\lemmaaa{Let $f$ a 0-evaluation faithful with respect to $S$. 
Then there exists a 0-evaluation $f'$, $PP_x$-similar to $f$ and faithful  with respect to $S'$.
}

\noindent
{\em Proof:~} 
First we of course define  $f'(i,R) = f(i,R)$  for each atom $ R\neq PP_x$, and each position $i$ in $R$, 
so the first condition of Definition \ref{podobne} is satisfied.

We will now define $f'(PP_x)$. Then we will notice that the second and third 
conditions  from Definition \ref{podobne} hold,
so $f'$ is indeed a 0-evaluation. Finally we will see that $f'$ is faithful with respect to $S'$.

Let  ${\cal I}(PP_x)=\{i_1, i_2\ldots i_s\}$, where ${\cal I}(PP_x)$ is the set of maximal important non-root  positions, as
in Definition \ref{skomplikowana}. 
Let $y_1, y_2,\ldots y_{s}$ be the important variables in positions $i_1, i_2\ldots i_s$ in $PP_x$ 
(the variables may repeat, this does not bother us).
For each $1\leq j\leq s$ let $d_j=f(y_j)$ (notice that this definition makes sense, because $y_j\in S$ for each $j$) 
and let $b_j=f(i_j,PP_x)$.


 Clearly, since $f$ is an evaluation, we have $b_j\equiv_0 d_j$ for all $j$.
But it means that we are now in the situation of Lemma About the Future (Lemma \ref{future1}), 
where $A=f(PP_x)$. 

So let $C$  be
as in  Lemma \ref{future1}. 
For  any  position $j\in top.pos(PP_x) $  define $f'(j,PP_x)$ as $C(j)$. 
Notice, that 
we can be sure (thanks to Lemma \ref{future1}) that $f'(j,PP_x) \equiv_0 f(j,PP_x)$.

Let now $j$ be a position in $PP_x$ which is not in $top.pos(PP_x)$. That means that the variable 
$z=PP_x(j)$ is in $S$.
Define $f'(j,PP_x)$ as $f(z)$. The condition $f'(j,PP_x) \equiv_0  f(j,PP_x)$ now holds trivially, since 
$f$ was a 0-evaluation.

We defined a function $f'$, which satisfies the first and the third condition from Definition \ref{podobne}.
Now we need to check that $Chase \models f'(PP_x)$. We know that $Chase \models C$, so this part of proof 
would be finished if we could show that $f'(PP_x)=C$. Of course by the definition of $f'$
 the atoms $f'(PP_x)$ and $C$ have equal 
elements of $Chase$ in the root and in all the positions in the set $top.pos(PP_x)$.
But this is not that clear what happens in the remaining positions.
   Surprisingly, this is the crucial moment, the one we 
spent long pages preparing for. The full power of
the normal form and family patterns is going to be used in the next 8 lines:

Consider two positions in $PP_x$: ~ $i\in \{i_1, i_2\ldots i_s\}$ and $j<_F i$. Let $z=P(j)$ and let 
$y$ be the variable in position $i$. 
Since $y$ is important, its 
parenthood atom, $PP_y$, is in $\psi $.

Since $\psi $ is in the normal form, we know, by the Ideological Condition, 
 that $PP_y(\delta(i,j))=z$. Since we defined
$f'(j,PP_x)$ to be $f(z)$, we get  $f'(j,PP_x) = f(\delta(i,j), PP_y)$. What we want to show is that $f'(j,PP_x)=C(j)$.
But this now follows directly from Lemma \ref{staloscnazw}.

In order to finish the proof 
of the Lemma we still need to notice that $f'$ is $S'$-faithful.
The atoms described by Definition \ref{eva}  are now all the atoms that were already contained in $S$, and one 
new atom $PP_x$. 
If $PP_x(j)$ was in $S$ we defined  $f(j,PP_x )$ as $f(PP_x(j))$, so we did not spoil
anything. 
The only problem could be with 
the values assigned to positions in $PP_x$ with variables from $S'\setminus S$.
But, by the Technical Condition 
each of these variables occurs in $PP_x$ only once, so the condition from Definition \ref{eva} is trivially satisfied.\eop

\bibliographystyle{alpha}
\bibliography{research}

\end{document}